\documentclass[lettersize,journal]{IEEEtran}
\UseRawInputEncoding
\usepackage{amsmath,amsfonts}
\usepackage{algorithm}
\usepackage{array}
\usepackage{multirow}
\usepackage{textcomp}
\usepackage{stfloats}
\usepackage{url}
\usepackage{verbatim}
\usepackage{graphicx}
\usepackage{threeparttable}
\usepackage{subcaption}
\usepackage{cite}
\usepackage{algpseudocode}
\usepackage{upgreek}

\begin{document}
\bstctlcite{IEEEexample:BSTcontrol}

\title{An on-chip Pixel Processing Approach with 2.4$\upmu$s latency for Asynchronous Read-out of SPAD-based dToF Flash LiDARs}

\author{
Yiyang Liu,
Rongxuan Zhang,
Istvan Gyongy,
Alistair Gorman,
Sarrah M. Patanwala,\\
Filip Taneski and 
Robert K. Henderson, \\
The University of Edinburgh, EH9 3FF Edinburgh, U.K.

\thanks{This work was supported by UK Engineering and Physical Sciences Research Council (EPSRC) Single Photons - Expanding the Spectrum (SPEXS), under EPSRC Reference ID EP/S026428/1.}
}

\maketitle

\begin{abstract}
We propose a fully asynchronous peak detection approach for SPAD-based direct time-of-flight (dToF) flash LiDAR, enabling pixel-wise event-driven depth acquisition without global synchronization. By allowing pixels to independently report depth once a sufficient signal-to-noise ratio is achieved, the method reduces latency, mitigates motion blur, and increases effective frame rate compared to frame-based systems. The framework is validated under two hardware implementations: an offline 256$\times$128 SPAD array with PC based processing and a real-time FPGA proof-of-concept prototype with 2.4$\upmu$s latency for on-chip integration. Experiments demonstrate robust depth estimation, reflectivity reconstruction, and dynamic event-based representation under both static and dynamic conditions. The results confirm that asynchronous operation reduces redundant background data and computational load, while remaining tunable via simple hyperparameters. These findings establish a foundation for compact, low-latency, event-driven LiDAR architectures suited to robotics, autonomous driving, and consumer applications. In addition, we have derived a semi-closed-form solution for the detection probability of the raw-peak finding based LiDAR systems that could benefit both conventional frame-based and proposed asynchronous LiDAR systems.
\end{abstract}


\section{Introduction}
\IEEEPARstart{L}{ight} Detection and Ranging (LiDAR) systems have emerged as a critical sensing modality for a wide range of applications, including autonomous driving\cite{Rapp_SPM2020, Roriz_ITIPS2022}, robotics\cite{Iqbal_Robotics2020}, consumer products\cite{vogt2021comparison}, and 3D mapping\cite{Luetzenburg_Protocal2024}, owing to their ability to provide high-resolution depth information with centimeter-level accuracy. Conventional LiDAR architecture typically employs a mechanical\cite{Zhou_IEEESensors2023} or Micro Electromechanical System\cite{Wang_Micromachines2020} (MEMS) based scanning mechanism to sequentially steer a narrow laser beam across the scene, capturing time-of-flight (ToF) information at each point. While scanning LiDAR offers high spatial resolution and long-range detection, it suffers from several intrinsic limitations, such as limited frame rate due to sequential acquisition, mechanical complexity, and susceptibility to motion artifacts, limiting its applicability to dynamic scenes or compact form-factor integration. 

\begin{figure}[t]
    \centering
    \includegraphics[width=\linewidth]{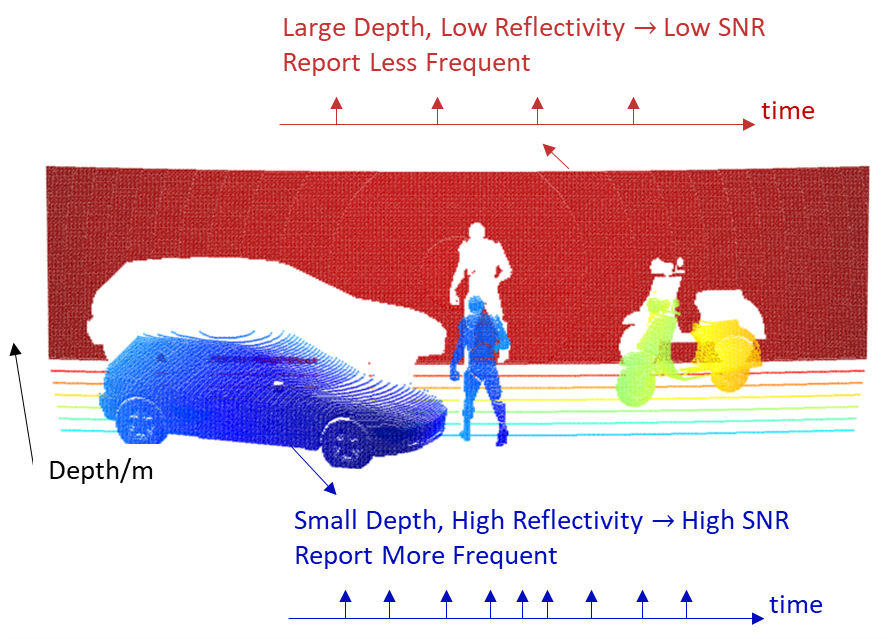}
    \caption{Illustration of the proposed asynchronous operation on two pixels with different SNR in a LiDAR point cloud.}
    \label{fig.fig1}
\end{figure}

Recent advances in single-photon detection and integrated circuit technologies have paved the way for flash LiDAR systems\cite{Niclass_JSSC2014, Sony_LiDAR_ISSCC2021, Sony_LiDAR_VLSI2025}, which eliminate the need for mechanical scanning through illuminating the entire scene in a single pulse and capturing the reflected signals simultaneously across a two-dimensional detector array. A key enabler of this architecture is the CMOS-compatible Single-Photon Avalanche Diodes(SPADs)\cite{Charbon_ICECS_Review}, which delivers single-photon sensitivity, picosecond-level timing accuracy, and scalability to high-resolution arrays\cite{Canon3_2MP, Canon2_1MP}. In contrast to conventional photodetectors such as Avalanche Photodiodes (APDs), SPADs can be tightly integrated with on-chip time-to-digital converters (TDCs) and histograms, allowing each pixel to independently measure time-of-flight with high temporal precision\cite{AhmetJSSC2019}. These advances in sensing and integration make fully solid-state, high-speed flash LiDAR systems feasible, offering real-time depth acquisition with improved robustness to motion and suitability for compact, high-throughput applications. These advances are further enhanced with the development of 3D stacking technology \cite{Tarek_3Dstack, Pavia_3Dstack}, where Back Side Illuminated SPAD pixels are on the top wafer \cite{Maciej_SPAD_Review}, and the processing circuit, which occupies a large area and decreases the fill factor of the sensor, on the bottom wafer that is usually designed in more advanced technology nodes \cite{hutchings2019reconfigurable, Preethi_2021_ICCSS}. More recently, with the integration of the on-chip peak searching methods for both full-histogram \cite{stoppa2021reconfigurable, Augusto_stacked} and partial-histogram \cite{Park_JSSC, Chao_JSSC, Adaps_OJSSCS} sensors, researchers have significantly decreased the output data rate and further enhanced the detection precision.

In the flash LiDAR systems, to maintain the accuracy of the measurement results in each frame, thousands of laser pulses are fired sequentially in each exposure to increase the Signal to Noise Ratio (SNR) of the resulting histogram. Therefore, the number of laser cycles within each frame is set to sustain a sufficient SNR according to the worst-case scenario - 10\% reflectance target at the nominal maximum detection distance. However, a larger number of laser cycles results in longer exposure time and reduced frame rate, making the system more susceptible to motion blur, especially in dynamic environments. In practice, only a small fraction of pixels within a typical scene experience such worst-case conditions. This means that the majority of the pixels have accumulated a histogram with enough SNR prior to those pixels with a lower reflected signal level, and they have to wait for them until the end of the exposure. Therefore, in this paper, we present a method to improve flash LiDAR performance in dynamic environments by allowing each pixel to operate asynchronously, which outputs depth data and resets immediately once a histogram reaches a sufficient SNR. As a result, the pixels with larger SNR do not need to wait for the global synchronous control signals, therefore report more often than the pixels with low SNR, as shown in Fig. \ref{fig.fig1}

The concept of asynchronous sensing has been explored in both Complementary Metal-Oxide-Semiconductor (CMOS) and Single-Photon Avalanche Diode (SPAD) based image sensors. In Time-to-First-Spike (TTFS) sensors \cite{Shoushun_TTFS}, each pixel integrates incoming signal until its accumulated charge surpasses a predefined threshold, triggering an event and initiating a reset. In dynamic vision sensors (DVS) \cite{Lichtsteiner_DVS}, pixels compare the current signal against a dynamically updated threshold, which is determined by the intensity level at the time of the last event, to decide whether to trigger a new output. With the advancement in the development of the SPAD-based pixels that naturally produce discrete spiking outputs, A. Berkovich et al., and C. Niclass et al. implemented asynchronous readout mechanisms directly in the readout circuits of SPAD arrays \cite{Berkovich20x20AER, Niclass64x48AER}. Building upon this, other works \cite{Sony_SPC_ISSCC2021} have combined the precise timing capability of SPADs with the TTFS concept to infer intensity information from photon arrival times. In addition, \cite{Lee_JSSC2025} and \cite{Francesco_DVS_SPAD} introduced in-pixel processing modules to emulate DVS-like functionality using digital logic within SPAD arrays. However, in both works, the overall readout architecture remained globally synchronized. Later, R. Gomez-Merchan et al. presented an asynchronous readout sensor with SPAD-based TTFS pixels, then converted the resulting data into DVS format offline using a Field Programmable Gate Array (FPGA) \cite{Gomez_SPADTTFS}. 

The works mentioned above primarily focus on asynchronous vision sensing, with some incorporating depth estimation capabilities by time gating the SPAD pixels and then reconstructing the time windows into a histogram off-chip. Recently, S. Park et al. presented a SPAD-based depth sensor capable of high frame rate operation \cite{Park_ISSCC2025}. In this system, each pixel independently validates its depth measurement by comparing the signal level against a dynamically estimated background noise threshold. This enables pixel-wise asynchronous acquisition with conditionally triggered, synchronized readout, resulting in spatially varying effective frame rates depending on object distance and reflectivity. However, since the sensor still relies on globally synchronized control signals for state transitions and data output, it limits the potential for fully asynchronous operation and retains a partially frame-based output structure. Furthermore, the background estimation is performed over a brief sampling window, which may occasionally lead to inaccurate thresholds under fluctuating ambient light or shot noise conditions. 

In this work, we present a fully asynchronous peak detection approach for direct time-of-flight (dToF) flash LiDAR systems. The proposed method is first demonstrated on a 256$\times$128 SPAD array-based LiDAR prototype \cite{HSLiDAR}. The asynchronous peak detection algorithm operates on the reconstructed histogram stream to emulate the full exposure process, enabling event-triggered depth acquisition. Based on this setup, we evaluate both the precision and accuracy of the proposed asynchronous LiDAR approach under various operating conditions. In addition, we introduce a method for reflectivity reconstruction by estimating the event rate at each pixel. Furthermore, we present a potential on-chip, in-pixel implementation solution of the proposed framework, and demonstrate the solution through an FPGA prototype. In this prototype, a pipelined processing is achieved with a novel background estimation and threshold computation method. Finally, we explore a DVS-inspired event representation by encoding depth changes between consecutive events, where positive and negative events correspond to increases and decreases in detected depth, respectively.

This paper is organized as follows. Section~\ref{Sec.2} presents the proposed asynchronous peak detection algorithm. In Section~\ref{Sec.3}, the architecture of the LiDAR system under test, the operation of the SPAD sensor, together with the experimental results in both static and dynamic conditions, are provided. Section~\ref{Sec.4} introduces a potential on-chip implementation of the proposed method, including a proof-of-concept FPGA prototype. Conclusions are drawn in Section~\ref{Sec.5}.

\vspace{-0.5em}
\section{Asynchronous Peak Detection Approach \label{Sec.2}}

\subsection{Peak Finding and Thresholding\label{sec.pk and thr}}
In flash LiDAR systems, the illuminator emits laser pulses across the entire scene at fixed intervals, with each repetition referred to as a laser cycle. Photons reflected from objects in the scene are detected by SPAD pixels, timestamped by time-to-digital converters (TDCs), and accumulated into per-pixel histograms. The background photon level in each bin during each laser cycle is assumed to be constant and is denoted by $B$. Due to the inherent nature of photon shot noise, the number of background photons detected in a given histogram bin $i$ follows a Poisson distribution, i.e., $b_i \sim \text{Poisson}(B)$, assuming the use of perfect multi-event TDCs (METDC) \cite{METDC}.

Over $N$ independent laser cycles, the histogram is formed by summing the photon arrivals across cycles. Since flash LiDAR systems typically include an idle period between the stop of histogram acquisition, and the start of next laser emission of consecutive laser cycles, the SPADs are fully recharged and ready to detect new photons at the start of each cycle, eliminating inter-cycle dead time effects. As a result, photon detections in different laser cycles are independent. According to the additivity property of the Poisson distribution \cite{Poisson_additive}, the total background count in bin $i$ after $N$ cycles is $b_i \sim \text{Poisson}(\lambda_b)$, where $\lambda_b=NB$. On the other hand, assuming the laser pulses have a Gaussian shape with a Full Width Half Maximum (FWHM) of $2\sqrt{2\ln 2}\sigma$, the returned signal from the object per laser cycle $S$ is expected to be
\begin{equation}
    S = \sum_{k=0}^{K} A_k f(t;\mu_k,\sigma), 
    \label{eq.1}
\end{equation}
where $K$ refers to the number of objects; $A_k$ is the number of signal photons detected by the pixel per laser cycle after attenuation, which is related to the Photon Detection Efficiency (PDE) of the SPADs, pixel area, emitted laser power, and various optical factors \cite{FBK_Model}; $\mu_k$ is the time of detection for the center of the Gaussian signal for the $k$th object; while $f(x)$ is the Probability Density Function (PDF) of the Gaussian distribution:
\begin{equation}
     f(x)=\frac{1}{\sqrt{2\pi}\sigma}e^{-\frac{(x-\mu)^2}{2\sigma^2}}.
     \label{eq.2}
 \end{equation} 
 
After discretization into histogram bins and considering photon shot noise, the photon counts belonging to the signal in a given bin $i$ are $h_{s,i} \sim \text{Poisson}(\lambda_{s,i})$, with
\begin{equation}
    \lambda_{s,i}=\sum_{k=0}^{k} A_kN \int_{t_{i}}^{t_{i+1}} f(t;\mu_k,\sigma)dt, 
    \label{eq.3}
\end{equation}
where $t_i$ is the starting time point for bin $i$. In this work, we define the peak as the bin with the largest photon count in the histogram, $i_{\max}$.

Once the peak bin of the histogram is determined, the next step is to assess whether the SNR of the histogram is sufficiently large for the peak to be considered significant. In LiDAR systems, the SNR of the histogram is defined as $NS/\sqrt{NB}$ \cite{Mora-Martin:24}. In this work, since the background noise is additive in the histogram rather than multiplicative, instead of defining a specific ratio-based SNR level, we use a threshold to separate background and signal bins. According to the Central Limit Theorem (CLT), when the number of laser cycles $N$ is sufficiently large, the Poisson distribution can be approximated by a normal distribution with both mean and variance equal to $\lambda$. Together with the additivity property of the Poisson distribution, the number of photons in bin $i$ follows:
\begin{equation}
    h_{i} \sim \mathcal{N}\left(\lambda_b+\lambda_{s,i}, \, \lambda_b+\lambda_{s,i}\right),
    \label{eq.4}
\end{equation}
and histogram bins that are only affected by background photons $h_b$ have $\lambda_{s,i} = 0$. Therefore, it is possible to state that any bins with a value of $\alpha$ standard deviation above the background mean are classified as a signal bin $h_s$. Hence, the threshold $I_{th}$ is set as
\begin{equation}
    I_{th}=\lambda_b+\alpha \sqrt{\lambda_b}.
    \label{eq.5}
\end{equation}

\subsection{Asynchronous Operation \label{sec.operation}}
\begin{figure}[htbp]
    \centering
    \includegraphics[width=0.95\linewidth]{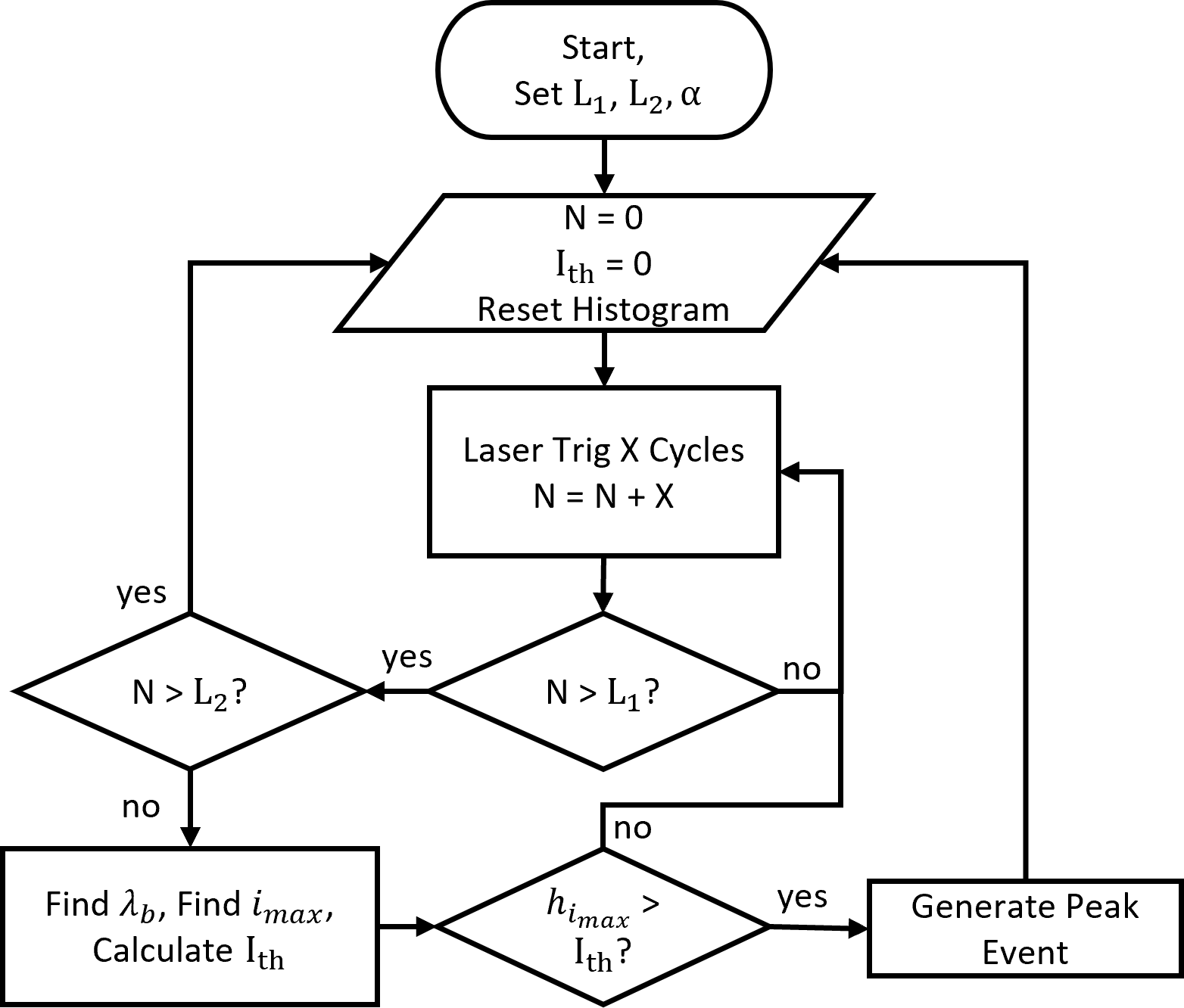}
    \caption{Pixel operation flow chart for the proposed asynchronous LiDAR.}
    \label{fig.operation}
\end{figure}

Following the definition of the histogram peak and threshold in the previous section, the overall operation of the proposed asynchronous LiDAR is shown in Fig.~\ref{fig.operation}. The process begins with the initialization of three hyperparameters: $L_1$, $L_2$, and $\alpha$. The LiDAR then repeatedly runs for $X$ laser cycles. After every $X$ cycles, the number of laser cycles $N$ is checked against two limiting parameters, $L_1$ and $L_2$. If $L_1<N<L_2$ after these $X$ cycles,  the background level of the current histogram $\lambda_b$ and peak bin $I_{\max}$ are computed. A threshold is then calculated and compared with $h_{i_{\max}}$ to determine whether there is a peak with sufficient SNR in the histogram. After a peak event is detected, or after the number of cycles exceeds the predefined upper limit $L_2$, the pixel histogram is reset in preparation for the next run.

\begin{figure}[htbp]
    \centering
    \includegraphics[width=\linewidth]{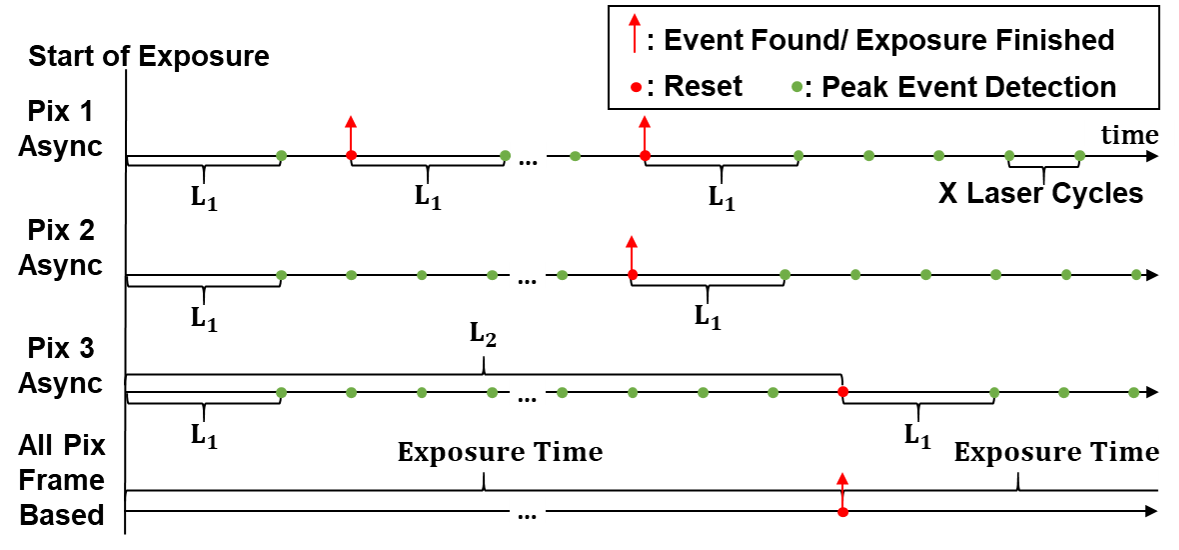}
    \caption{Example operation of different pixels.}
    \label{fig.pix operation}
\end{figure}

The operation starts concurrently across all pixels but proceeds asynchronously, causing each pixel to reset at different times, as shown in Fig.~\ref{fig.pix operation}. In this figure, pixels 1, 2, and 3 are distinct pixels with decreasing SNR from pixel 1 to pixel 3. As a result, pixel 1 reports a peak event and resets soon after $L_1$, followed by pixel 2 reporting at a later stage. Since each pixel resets its histogram after detecting a peak event, it immediately begins forming a new histogram for the next detection cycle. Consequently, given sufficient SNR, pixel 1 is able to report a second event. Pixel 3 illustrates a low-SNR case when no significant peak is found within $L_2$ limit. Therefore, it is forcibly reset at $L_2$ to prevent histogram saturation. In contrast, all pixels in conventional frame-based systems operate under a fixed exposure interval, no matter how SNR varies, as indicated in the bottom row of Fig.~\ref{fig.pix operation}.

In practice, the value of three hyperparameters needs to be determined prior to the start of the exposure. The value of $L_2$ is set equal to the number of laser cycles typically used as the exposure time in frame-based LiDAR systems. This choice is based on a system-level parameter that ensures, under nominal working conditions, all pixels are expected to accumulate histograms with sufficient SNR within that time frame. The selection of the other two parameters is discussed in Section~\ref{sec.parameterseletion}.

\subsection{Analysis of System Performance \label{sec.parameterseletion}}

During actual measurements, due to photon shot noise, the largest bin may not correspond to the center of the returned signal in the histogram $i^*$, where the signal photon count is
\begin{equation}
    h_{i^*}=NA\int_{t_{i^*}}^{t_{i^*+1}}f(t;\mu,\sigma)dt,
    \label{eq.6}
\end{equation}
where $t_{i^*}$ is the starting time point of the peak bin. Also, the threshold set in equation Eq.~\ref{eq.5} could potentially cause a false positive if a background bin value is too large. In this section, we analyze the True Positive Rate (TPR) $P_{true}$ and false positive rate (FPR) $P_{false}$ of the proposed peak finding and thresholding method and derive a method to set an optimal value for the hyperparameter $\alpha$. For simplicity, we assume there is only one reflection in the measurement and that the photon counts in the bins are i.i.d. (independent and identically distributed). Although neither assumption is strictly true in real-world systems due to multi-path effects \cite{AfterPulse} and SPAD secondary effects such as dead time and after-pulsing \cite{MultiPath}, they provide a sufficiently accurate theoretical model for analysis purposes.

\subsubsection{Accuracy Analysis}
\begin{figure}[htbp]
    \centering
    \includegraphics[width=\linewidth]{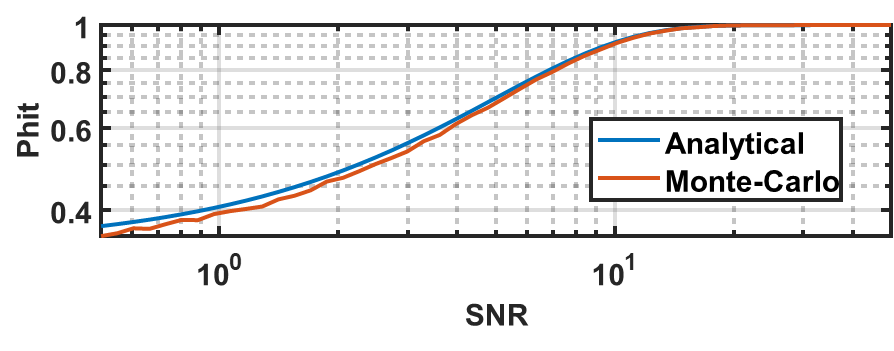}
    \caption{Monte-Carlo validation of Eq.~\ref{eq.7}, with $M_B=7$, $B=0.05$, with $\mu$ positioned at the edge of histogram bin, and $\sigma=1.5\Delta t_i$.}
    \label{fig.Phit_vs_SNR_MC}
\end{figure}

\begin{figure}[htbp]
    \centering
    \includegraphics[width=\linewidth]{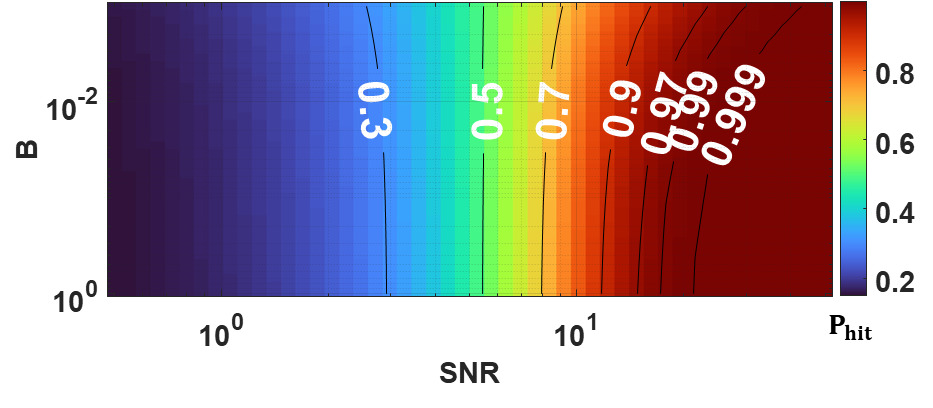}
    \caption{Variation of $P_{hit}$ under different $B$ and SNR, with $M_B=7$, $\mu$ at the edge of histogram bin, and $\sigma=1.5\Delta t_i$.}
    \label{fig.Phit_vs_SNR}
\end{figure}

First, we define the hit rate as the probability that the maximum bin corresponds to the true peak bin, i.e., $i_{\max}=i^*$. Under the assumption of independence, $P_{hit}$ satisfies:
\begin{align}
    P_{hit}&=P(h_{i^*}>\max_j h_j)\nonumber\\
        &=\int^\infty_0 P(\max_j h_j<x)\cdot f_X(x) dx,
    \label{eq.7}
\end{align}
where $f_X(x)$ is the PDF of the Gaussian-distributed $h_{i^*}$ (see Eq.~\ref{eq.2}), with $\mu=\sigma^2=\lambda_{s,i^*}+\lambda_b$; $P(\max_j h_j<x)=\Phi((x-\lambda_b)/\sqrt{\lambda_b})^{M_B}$; $\Phi(x)$ is the Gaussian Cumulative Distribution Function (CDF); and $M_B$ is the number of background bins. In this analysis, we ignore the probability of the maximum bin being another signal bin for two reasons: first, in systems using raw peak-finding methods without filtering, the optimized $\sigma$ of the laser pulse is 1 to 1.5 times the histogram bin \cite{koerner2021models}; second, this error can be corrected by sub-bin interpolation methods, e.g., quadratic interpolation \cite{padmanabhan2021depth}, or the Center of Mass Method (CMM) \cite{CMM}. Since we have assumed $K=1$, and the integral of the Gaussian PDF is 1, $SNR=NA/\sqrt{NB}$, the SNR can be enhanced by increasing the emitted laser power $A$, suppressing ambient noise $B$, or increasing the number of laser cycles $N$. To validate Eq.~\ref{eq.7}, a Monte-Carlo simulation is performed under the condition of $\sigma=1.5\Delta t_i$, with $\mu$ positioned at the edge of a histogram bin, and the result of the simulation is compared to the numerical result of Eq.~\ref{eq.7} as shown in Fig.\ref{fig.Phit_vs_SNR_MC}, where two curves match each other closely. Under the same system condition, the variation of $P_{hit}$ with respect to background level and histogram SNR is shown in Fig.~\ref{fig.Phit_vs_SNR}. According to this plot, to achieve a 99\% hit rate, the SNR must be set around 18. 


\subsubsection{Effect of $\alpha$}
We now examine how varying $\alpha$ impacts $P_{true}$. Defining the probability of the maximum bin exceeding the threshold as $P_{pass}$, we have:
\begin{align}
    P(pass|hit)&=P(h_{i^*}>I_{th})\nonumber\\
    &=1-\Phi\left(\frac{I_{th}-(\lambda_{s,i^*}+\lambda_b)}{\sqrt{\lambda_{s,i^*}+\lambda_b}}\right)\nonumber\\
    &=1-\Phi\left(\frac{\alpha\sqrt{\lambda_b}-\lambda_{s,i^*}}{\sqrt{\lambda_{s,i^*}+\lambda_b}}\right).
    \label{eq.8}
\end{align}
Based on Eq.~\ref{eq.7} and Eq.~\ref{eq.8}, $P_{true}=P_{hit}\cdot P(pass|hit)$. In a real LiDAR system, since a large $\alpha$ increases the detection threshold, thereby suppressing both $P(pass|hit)$ and $P_{true}$. Given a required $P_{true}$ and known system parameters, this allows for the derivation of an upper bound on $\alpha$, as shown in Eq.~\ref{eq.9}.
\begin{align}
    \alpha_{max}&=\frac{\lambda_{s,i^*}+\Phi^{-1}(1-P_{true}/P_{hit})\sqrt{\lambda_b+\lambda_{s,i^*}}}{\sqrt{\lambda_b}}\nonumber\\
    &=h_{i^*}\cdot SNR+ \Phi^{-1}\left(1-\frac{P_{true}}{P_{hit}}\right)\cdot\sqrt{1+\frac{h_{i^*}\cdot SNR}{\sqrt{\lambda_b}}}.
    \label{eq.9}
\end{align}

For the false positive rate, since the probability of a single bin being smaller than the threshold is $\Phi(\alpha)$, the total false positive rate is given by
\begin{equation}
    P_{false} = 1-\Phi(\alpha)^{M_B}. 
    \label{eq.10}
\end{equation}
Thus, the minimum possible value of $\alpha$ is derived as
\begin{equation}
    \alpha_{min} = \Phi^{-1}\left((1-P_{false})^{1/M_B}\right).
    \label{eq.11}
\end{equation}

\begin{figure}[htbp]
    \centering
    \includegraphics[width=\linewidth]{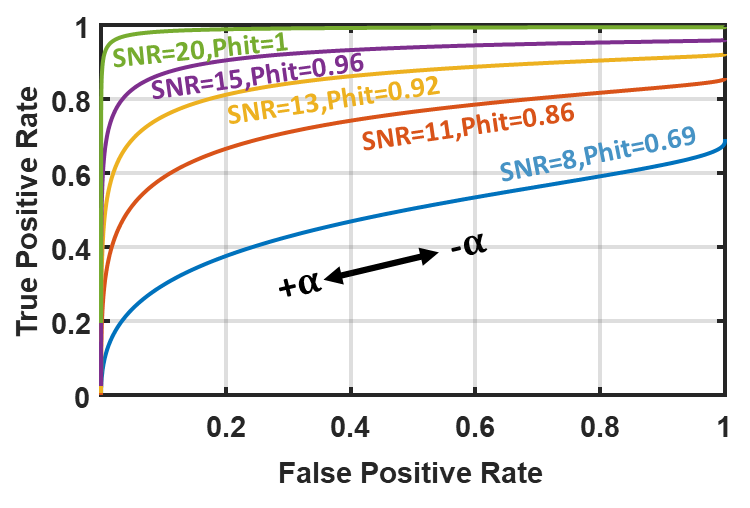}
    \caption{ROC plot of thresholding, by sweeping $\alpha$ and $SNR$, with $M_B=7$, $B=0.05$, $\mu$ at the edge of histogram bin, and $\sigma=1.5\Delta t_i$.}
    \label{fig.Phit_ROC} 
\end{figure}

Based on Eq.~\ref{eq.8} and Eq.~\ref{eq.10}, the Receiver Operating Characteristic (ROC) curve can be plotted to evaluate the performance of the thresholding method in Eq.~\ref{eq.5} on the raw peak finding LiDAR data, as shown in Fig.~\ref{fig.Phit_ROC}. In this figure, each ROC curve is generated by alternating the value of $\alpha$, where a larger $\alpha$ leads to a larger TPR but also increases the FPR. Additionally, the Area Under Curves (AUC) for each ROC curve is primarily influenced by the histogram SNR, as the maximum achievable TPR is constrained by $P_{hit}$, which is closely related to the SNR of the histogram.

Among all of the parameters, $P_{hit}$ reflects the intrinsic accuracy of the LiDAR system, determined solely by the system design and the number of laser cycles, and should ideally remain high. The term $P(pass|hit)$ governs how frequently peak events are generated. When $P_{hit}$ is sufficiently high, a relatively modest value of $P(pass|hit)$ can be tolerated, thereby relaxing the constraints on the choice of $\alpha$. However, reducing $P(pass|hit)$ may also decrease the event reporting rate, as peaks are less likely to exceed the threshold at a given SNR. Finally, the false positive probability $P_{false}$ depends solely on $\alpha$, increasing $\alpha$ can suppress false detections $P_{false}$ but at the cost of reducing $P(pass|hit)$. This trade-off must be carefully managed to optimize system performance.

\subsubsection{Effect of $L_1$}
In addition to $\alpha$, the value of $L_1$ will also affect the performance of the proposed asynchronous LiDAR. Specifically, $L_1$ defines the minimum number of laser cycles required before peak detection and thresholding can takes place. This serves as a key constraint to ensure statistical reliability, as required by the law of large numbers under the CLT. Based on the derivation in section~\ref{sec.parameterseletion}, the Poisson-distributed photon counts of the histogram bins are approximated as Gaussian-distributed. This statement only holds true when the expected photon counts $\lambda_b$ is sufficiently large.. As a rule of thumb, $L_1$ must be set to let $\lambda_{b,min}=L_1B\ge30$ \cite{Witte2021-fa}. In addition, $L_1$ constrains the initial SNR of the histogram by setting the minimum laser cycles $N$ required for evaluation. As such, it is closely related to the value of $P_{hit}$ of the histogram. Increasing $L_1$ improves the initial detection accuracy by allowing more photon counts to accumulate, which raises $P_{hit}$. However, if $L_1$ is set too high, the minimum interval between each two consecutive peak events is also increased. This results in higher latency, resembling the behavior of traditional frame-based systems.

\section{Experiments\label{Sec.3}}
\subsection{System Under Test}
\begin{figure}[htbp]
    \centering
    \includegraphics[width=\linewidth]{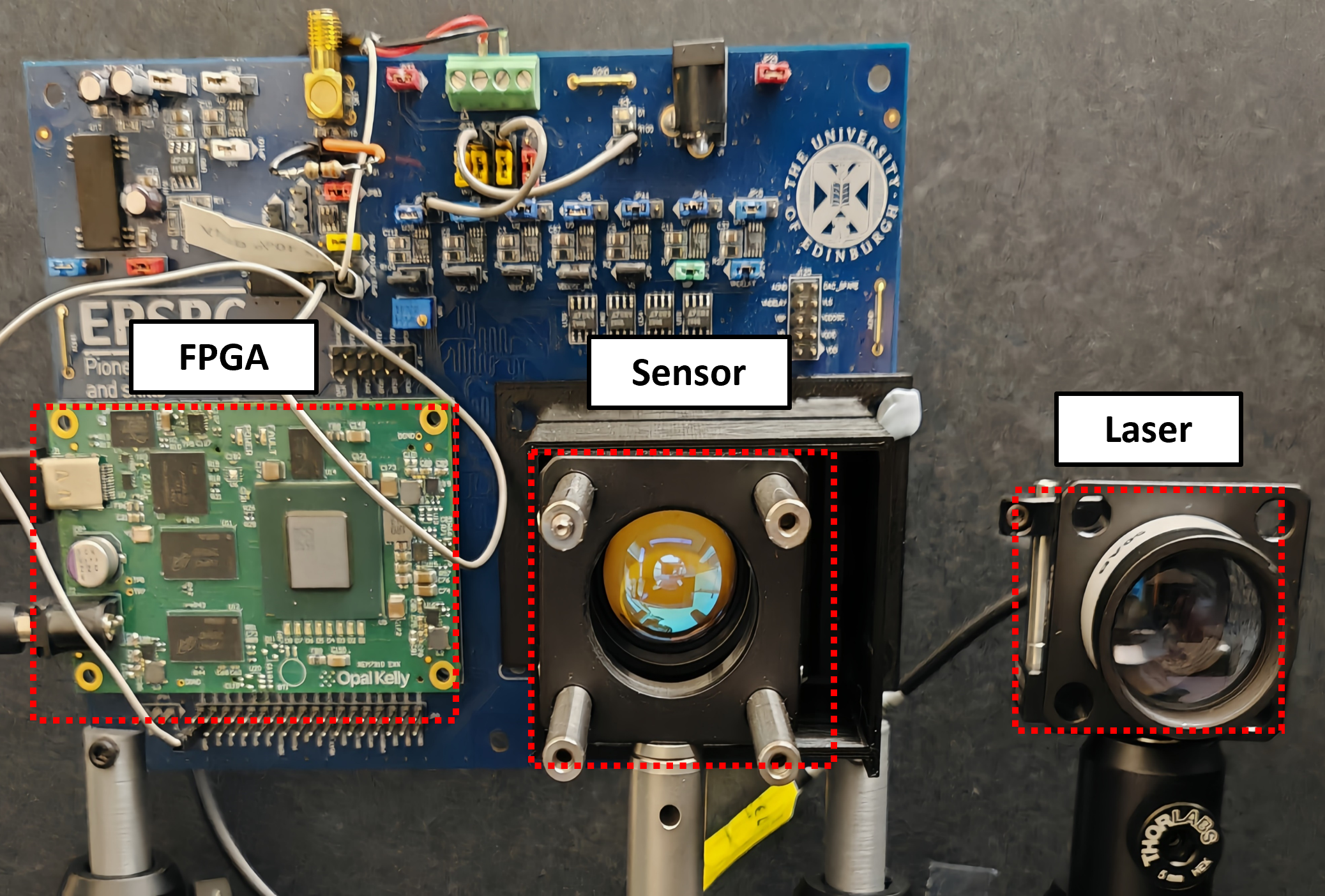}
    \caption{Experimental setup.}
    \label{fig.HSLiDAR_System} 
\end{figure}

To evaluate the feasibility of the proposed asynchronous LiDAR system, we set up a prototype system, as illustrated in Fig.~\ref{fig.HSLiDAR_System}. The system consists of a SPAD-based image sensor, a flash laser diode, and an FPGA that controls the operation and provides the interface with a PC.

\begin{figure*}[htbp]
    \centering
    \includegraphics[width=\linewidth]{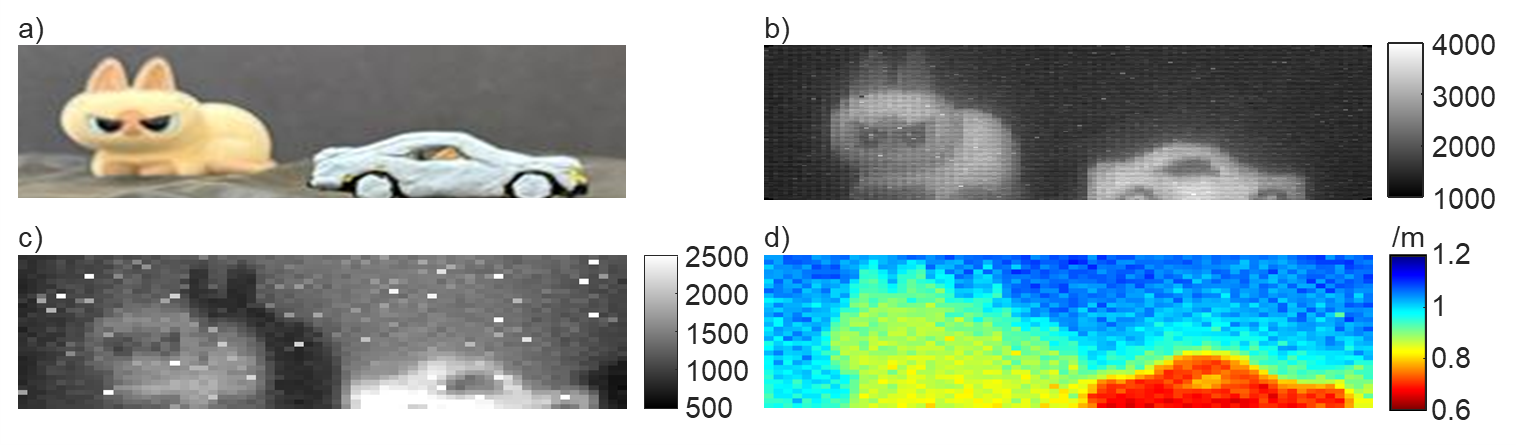}
    \caption{Static scene used in~\ref{sec.static}: (a) RGB image; (b) passive photon-counting image with the laser off; (c) active photon-counting image with the laser on and filter applied; (d) depth map obtained from the frame-based model.}
    \label{fig.static_scene} 
\end{figure*}

The SPAD image sensor used in this setup is described in \cite{HSLiDAR}. Fabricated in 40 nm FSI technology, it contains 256$\times$128 passively quenched SPADs, with each group of 4$\times$4 neighboring SPADs combined to form a macropixel. Although the sensor integrates an internal per-pixel gated ring oscillator-based METDC, in this work, we employ an external 114 MHz clock to define the bin width of the TDC and, consequently, the histogram. This configuration results in a bin width of 8.75 ns. In addition, the sensor supports the sliding partial histogram method \cite{taneski2022laser}, which provides 128 windows with 8 bins each. To emulate the behavior of the asynchronous LiDAR under full-histogram operation, we fix the time gate to the window containing the peak of the returned signal. Consequently, the system’s histogram output is limited to 8 bins.

The sensor is connected with an Opal Kelly XEM6310 FPGA, which provides the sensor firmware, the external timing clock, and generates a 1.2 MHz, 1.2\% duty-cycle trigger signal to drive the 2 W peak-power OSELA 860 nm TOFI laser module. Since the sensor architecture is frame-based, we configure it to operate with the minimum possible exposure time, producing sequential histograms containing photons from only 5 laser cycles. These consecutive frame-based histograms serve as the input data for testing the proposed asynchronous approach. During post-processing, the Center of Mass Method (CMM) is applied to estimate the sub-bin position of the peak and thus improve depth accuracy. Finally, the detected peak events are reported together with their corresponding depth information.

\subsection{Static Scene \label{sec.static}}

For the first experiment, objects (a cat and a car) were placed in the imaging scene. The passive photon count, active photon count, and depth map acquired in frame mode over 2000 laser cycles, together with the RGB image for reference, are shown in Fig.~\ref{fig.static_scene}. Note that the active photon count and depth map images were generated with pixels combined into macropixels, reducing the resolution by a factor of 16.  

\begin{figure}[htbp]
    \centering
    \includegraphics[width=\linewidth]{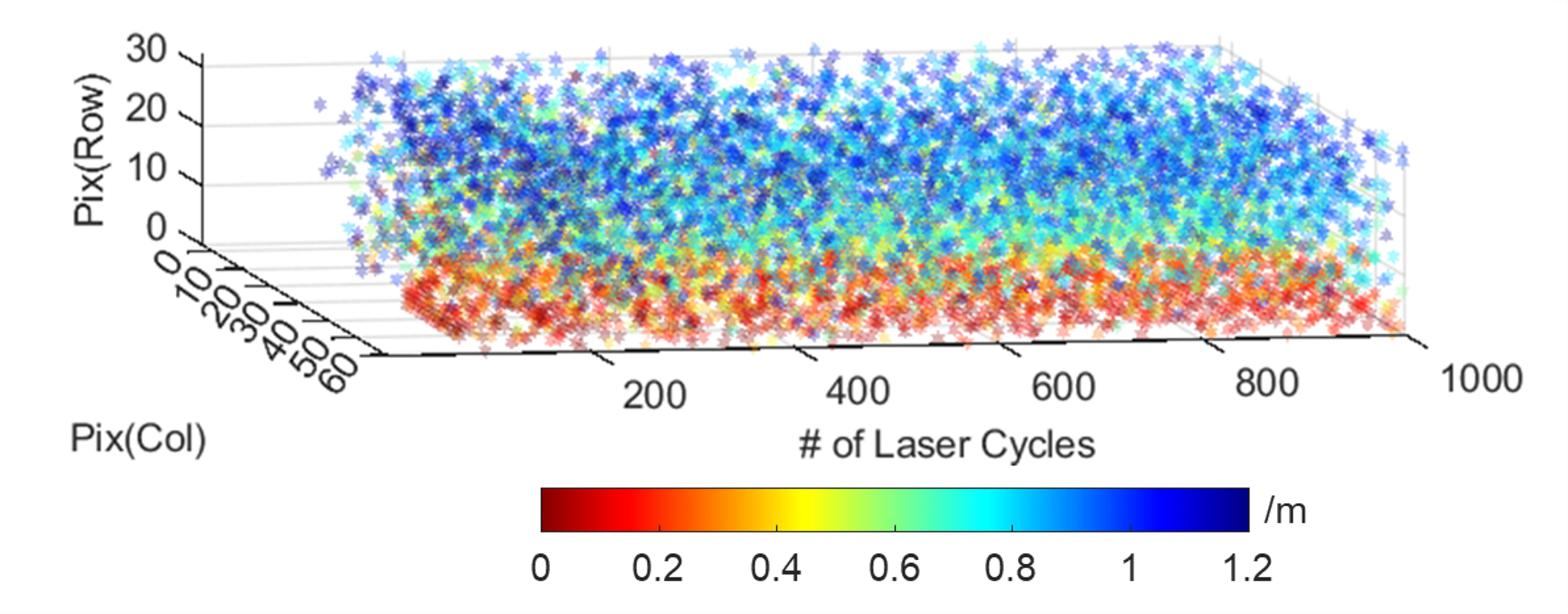}
    \caption{Spatio-temporal event plot of the static scene under $\alpha$ = 8, $L_1$ = 100, $L_2$ = 2000, and $X$ = 10.}
    \label{fig.statc_Async} 
\end{figure}

\begin{figure}[htbp]
    \centering
    \includegraphics[width=\linewidth]{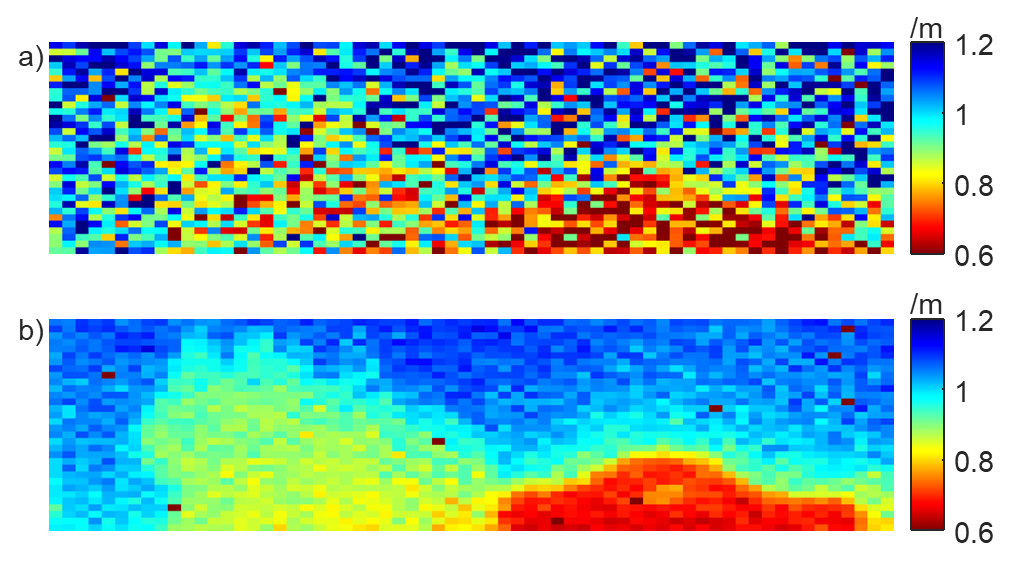}
    \caption{Depth map of (a) the final peak event, and (b) the average of the last 10 peak events.}
    \label{fig.final_event} 
\end{figure}

\begin{figure}[htbp]
    \centering
    \includegraphics[width=\linewidth]{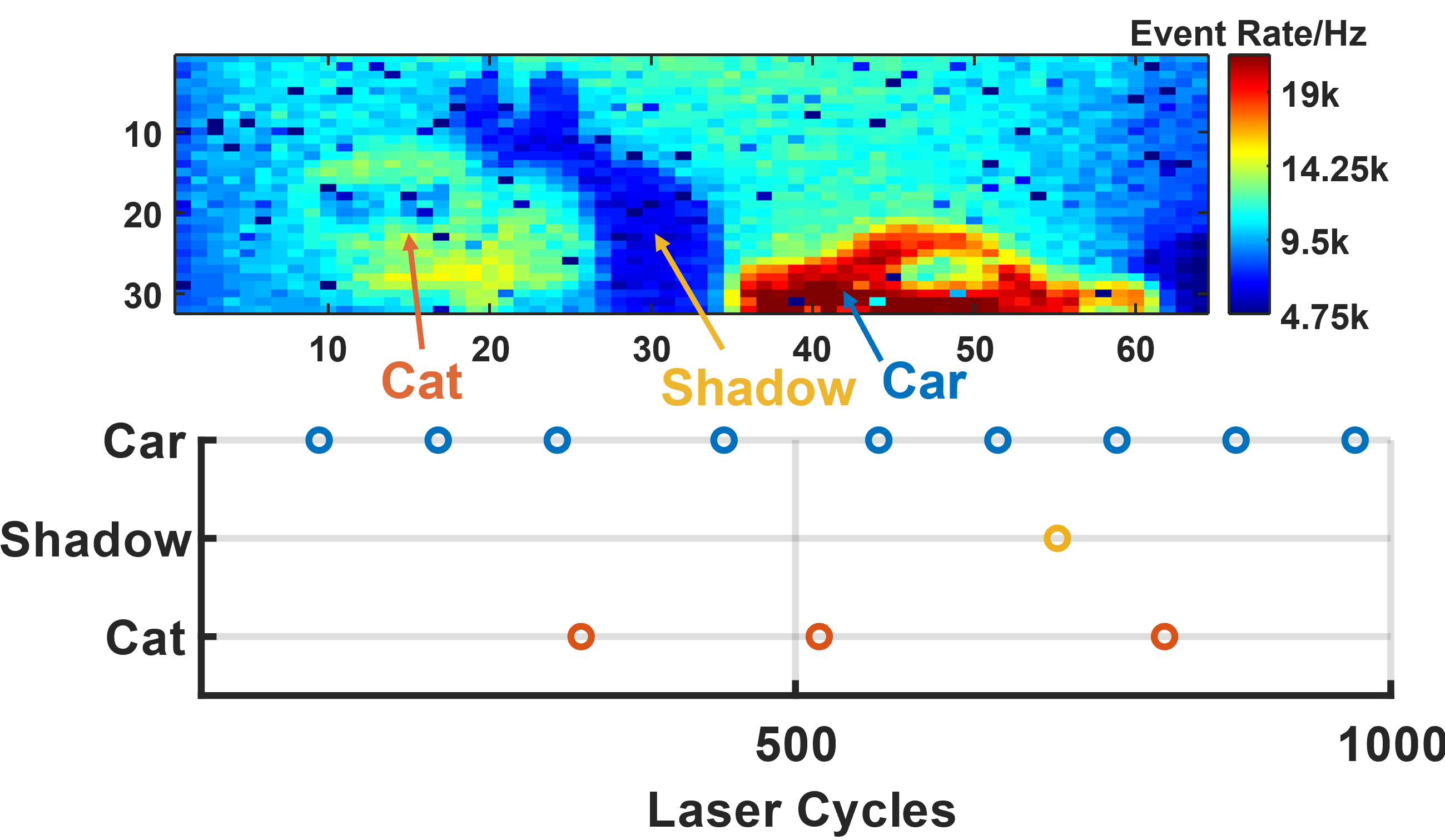}
    \caption{Counts of the events in each pixel, and event spikes for selected pixels from different objects.}
    \label{fig.event_counts} 
\end{figure}

A dataset of 10,000 laser cycles (2000 five-cycle frames) was recorded and processed using the PC-based asynchronous system emulator with parameters $\alpha$ = 8, $L_1$ = 100, $L_2$ = 2000, and $X$ = 10. The resulting spatio-temporal event plot is shown in Fig.~\ref{fig.statc_Async}. Each point corresponds to a detected peak event at a specific time and pixel, with the color indicating the depth after CMM interpolation. Distinct objects are clearly separated by color: blue points correspond to the background wall, green/yellow to the cat, and red to the car. In this plot, pixels begin exposing at different times due to varying reflected signal strengths, which lead to different SNR values across pixels. Pixels corresponding to high-SNR regions (e.g., the car) generate denser event clouds at earlier stages (e.g., within the 0--200 cycles interval). In contrast, objects located farther away (e.g., the background and the cat) exhibit sparser event clouds during the same period. This observation confirms the effectiveness of the proposed asynchronous operation. In addition, this plot shows a clear empty region before 100 laser cycle, this is due to the selection of $L_1$ = 100.

Compared with the frame-based depth map in Fig.~\ref{fig.static_scene}(d), the single-event results in Fig.~\ref{fig.statc_Async} appear noisier. To highlight this difference, the last peak events of each pixel were combined into a depth map, shown in Fig.~\ref{fig.final_event}(a). This degradation arises because the asynchronous method only compares the maximum bin of the histogram against the threshold, whereas sub-bin interpolation also depends on signal photons in adjacent bins. These adjacent bins may not have accumulated sufficient photons within the limited laser cycles. To mitigate this effect, an averaging process over 10 consecutive peak events was applied, producing the depth map shown in Fig.~\ref{fig.final_event}(b). This result is noticeably less noisy and comparable to the frame-based output in Fig.~\ref{fig.static_scene}(d). The dark-red hot pixels indicate locations where no events were detected within 10,000 cycles.  

The distribution of peak event counts across all pixels is presented in Fig.~\ref{fig.event_counts}, showing a clear correlation with object depth and reflectivity. Pixels corresponding to closer or more reflective surfaces generate higher photon counts and therefore more frequent peak events. The timing of events for three representative pixels—on the car, the cat, and the cat’s shadow—further demonstrates asynchronous operation, as detections occur asynchronously at different times for each pixel. Moreover, the frequency of event reporting aligns well with the laser signal strength by referring back to Fig.~\ref{fig.static_scene}(c).  

\begin{figure}[h]
    \centering
    \includegraphics[width=\linewidth]{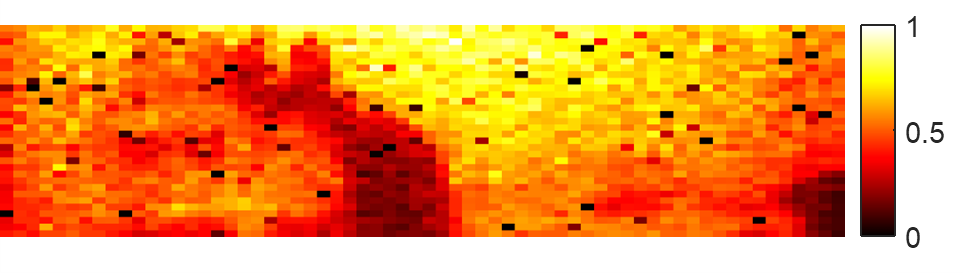}
    \caption{Reconstructed relative reflectivity of the scene.}
    \label{fig.reflectivity} 
\end{figure}

Beyond depth information, LiDAR systems also provide the relative reflectivity $\rho$ of objects in the scene \cite{hata2014road}, typically calculated as $\rho = P_{rx} \times z^2$, where $P_{rx}$ is the received photon count and $z$ is object depth. Since $P_{rx}$ is directly proportional to $\rho$ and inversely proportional to $z^2$ \cite{FBK_Model}, reflectivity can be inferred accordingly. In the proposed asynchronous approach, $P_{rx}$ is replaced with the number of peak events. The reconstructed normalized reflectivity is shown in Fig.~\ref{fig.reflectivity}.  

\subsection{Hyperparameter Analysis}
\begin{figure}[h]
    \centering
    \includegraphics[width=\linewidth]{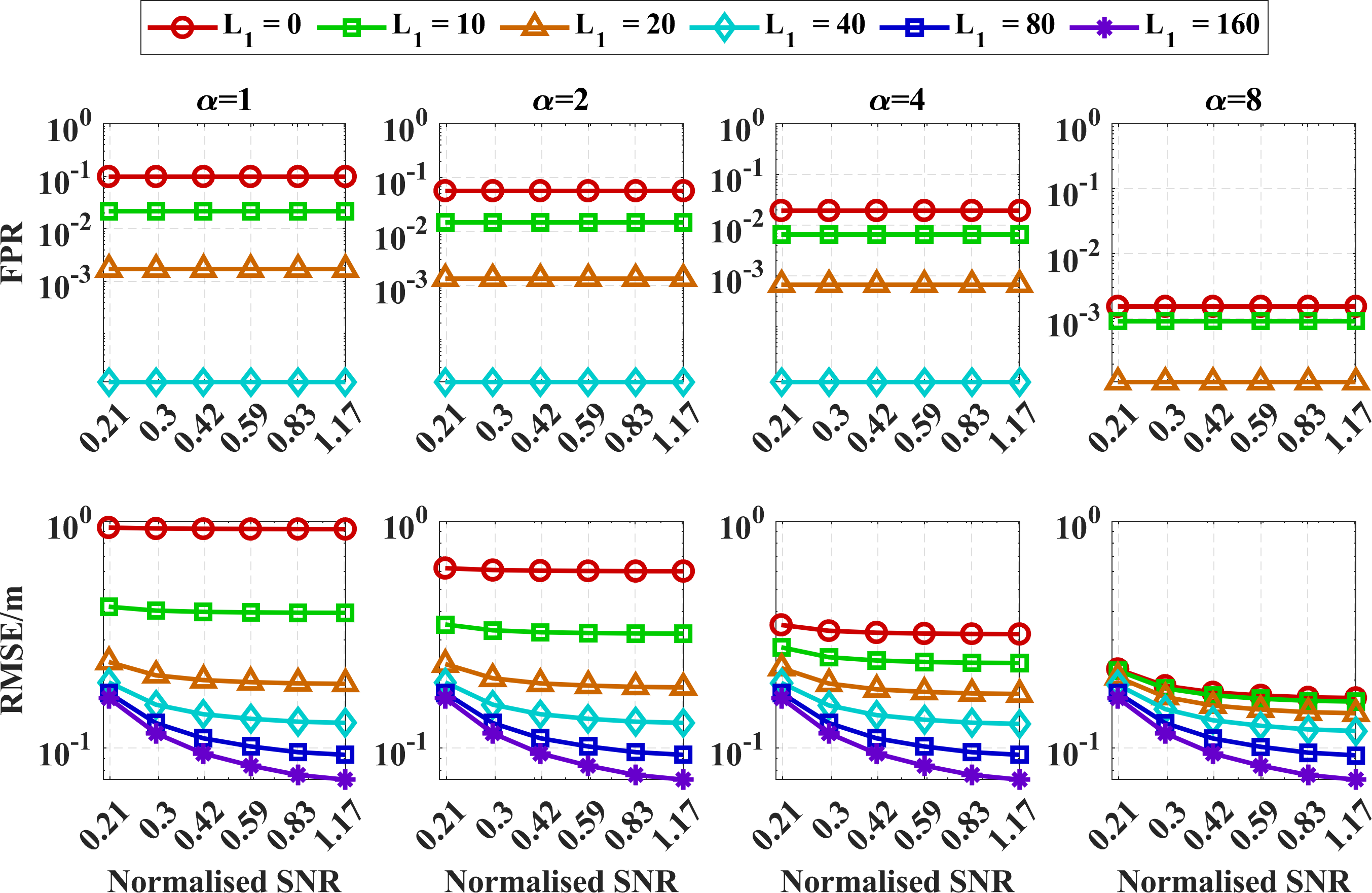}
    \caption{Analysis of the effect of hyperparameters.}
    \label{fig.paramana} 
\end{figure}

The next set of experiments evaluated the effect of varying the hyperparameters $\alpha$ and $L_1$ on detection accuracy. In this experiment, a flat board was placed 1.15 m away from the LiDAR system. At this position, the returned signal peak of most pixels lay on the edge of two histogram bins, creating a worst-case scenario for peak finding. By varying the SNR, $\alpha$, and $L_1$ of the asynchronous system, the false positive rate (FPR) and root mean square error (RMSE) of the detection were obtained, as shown in Fig.~\ref{fig.paramana}. In this plot, a false detection is defined as any peak event occurring outside of $\pm0.5$ bin of the histogram after CMM interpolation, and the RMSE is calculated between the detected peak events and the ground-truth distance of 1.15 m. Note that the SNR in this plot is the normalized SNR, defined as the average SNR of the histogram in a single laser cycle $A/B$, and is independent of the number of laser cycles.

From Fig.~\ref{fig.paramana}, it is clear that the FPR is not related to the value of SNR, consistent with the mathematical analysis in Eq.~\ref{eq.10}. However, as SNR increases, the RMSE decreases when $L_1$ is larger than 20. Increasing $\alpha$ suppresses the FPR but has no effect on RMSE if $L_1 > 20$. Finally, larger values of $L_1$ improve both indicators, but the marginal benefit diminishes once $L_1$ exceeds 40.

Overall, based on these results and the analysis in Section~\ref{sec.parameterseletion}, for a proposed asynchronous system with fixed normalised SNR (as the optical setup is not interchangeable), the first parameter to determine is $L_1$. It must be chosen to satisfy the Gaussian approximation. In the system under test, $L_1$ should be set to 20 or 40: a smaller $L_1$ leads to poor detection accuracy, while a larger $L_1$ provides limited improvement and diminishes the advantage of asynchronous operation. The choice of $\alpha$ is solely dependent on the acceptable FPR, as $\alpha$ does not affect RMSE. For example, if the desired FPR is below 0.1\% with $L_1 = 20$, then $\alpha = 4$ should be selected.

\subsection{Dynamic Scene \label{sec.motion}}
The previous section discussed the performance of the proposed method under static objects, which corresponds to the operating conditions of most indoor depth cameras. However, static scenes do not fully benefit from asynchronous operation, since fast readout is unnecessary to avoid motion blur. In this section, two experimental scenes were set up to represent common types of motion in LiDAR detection: (1) \textit{radial motion}, where an object moves toward and away from the LiDAR system along the radial direction, and (2) \textit{transverse motion}, where an object moves horizontally in the image plane, perpendicular to the radial direction.

\begin{figure}[htbp]
    \centering
    \includegraphics[width=\linewidth]{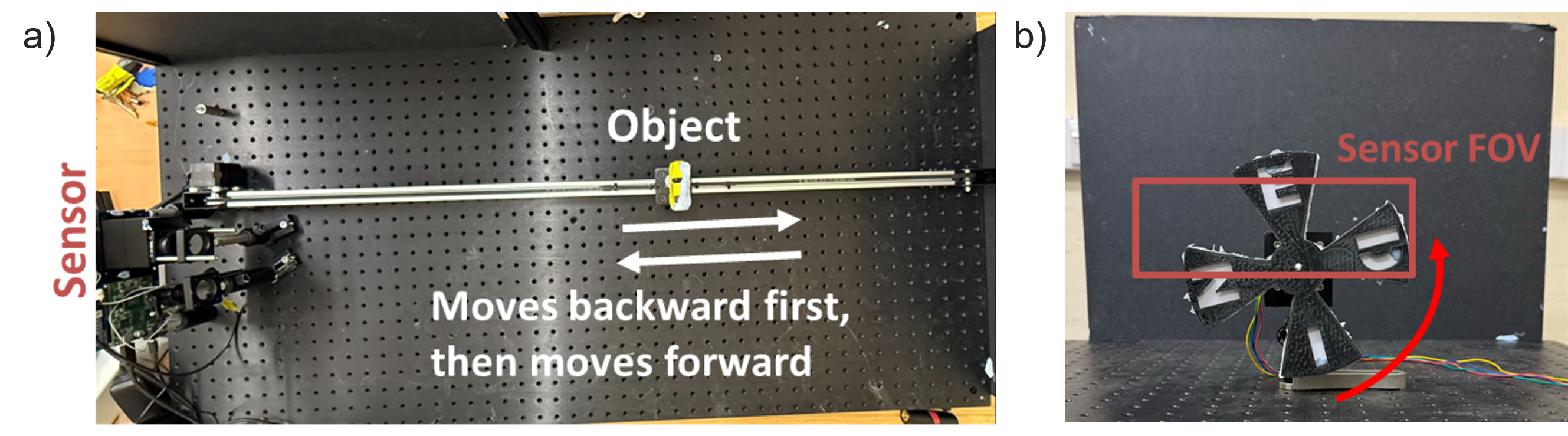}
    \caption{Experimental scenes: (a) radial motion, (b) transverse motion.}
    \label{fig.motion_scene} 
\end{figure}

\begin{figure}[htbp]
    \centering
    \includegraphics[width=\linewidth]{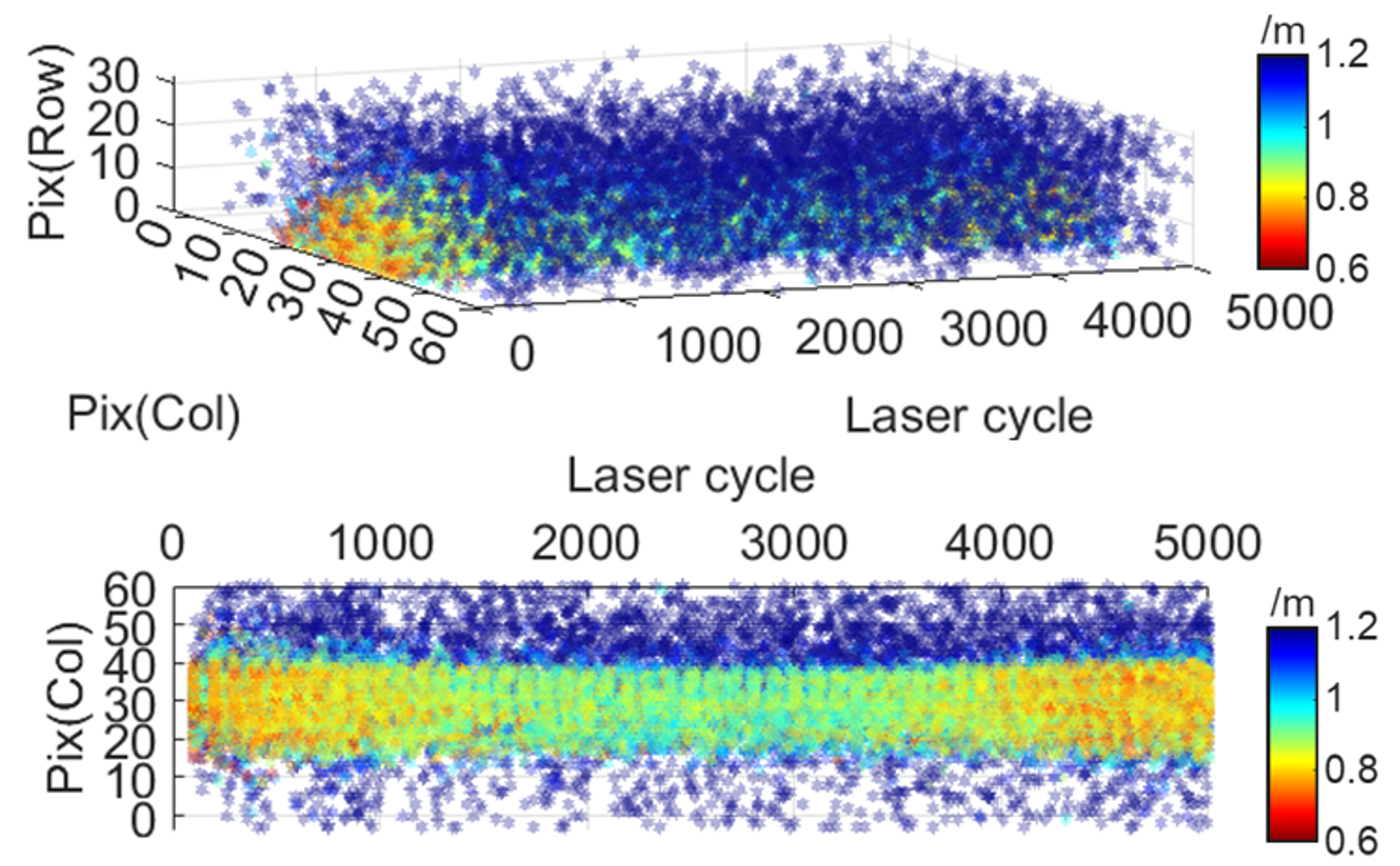}
    \caption{Event point cloud of the radial motion with two different views, using $\alpha=8$ and $L_1=40$.}
    \label{fig.radial_motion_cloud} 
\end{figure}

The setups for radial and transverse motion are illustrated in Fig.~\ref{fig.motion_scene}. For the radial motion, a toy car was mounted on a linear stage controlled by a stepper motor and an Arduino microcontroller unit (MCU). The car was first moved backward away from the sensor and then returned to its original position. After processing the data with $\alpha=8$ and $L_1=40$, the resulting event point cloud is shown in Fig.~\ref{fig.radial_motion_cloud}. In the top plot, the shape of the car is clearly separated from the background wall. The bottom plot shows that the depth of the peak events follows the programmed motion, first increasing and then decreasing. Another observation is that the apparent size of the object decreases as it moves further away, which matches the physical reality. In addition, pixels corresponding to the background wall, which have lower SNR, report less frequently than car pixels, consistent with the static scene results in Fig.~\ref{fig.event_counts}.

\begin{figure}[htbp]
    \centering
    \includegraphics[width=\linewidth]{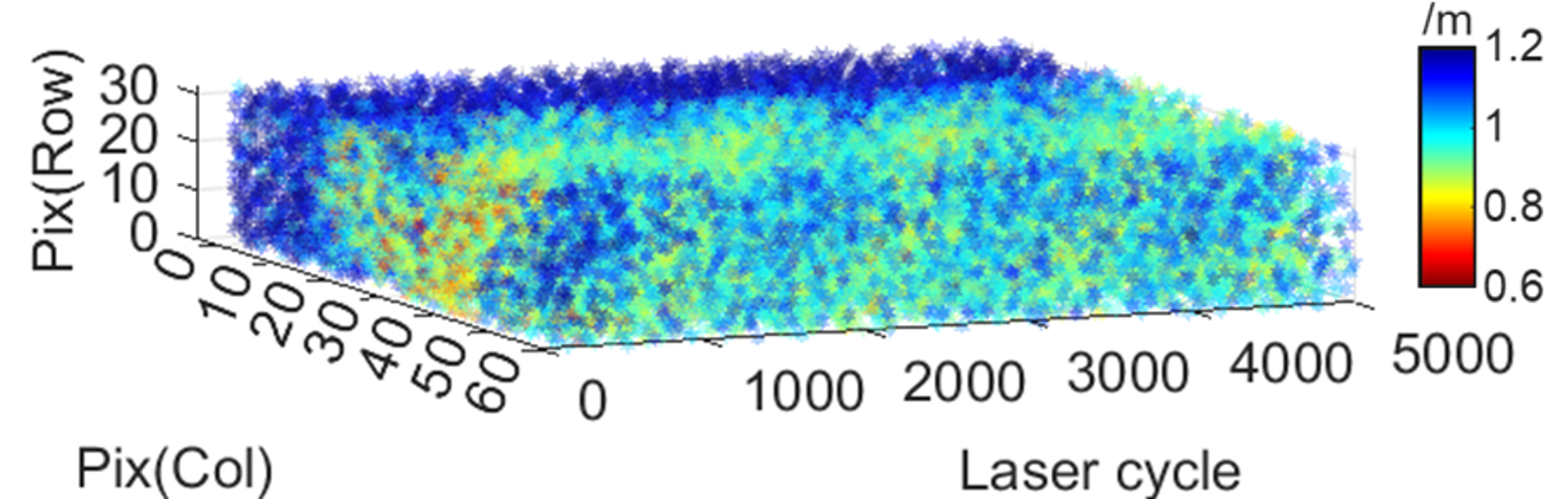}
    \caption{Event point cloud of the transverse motion, using $\alpha=8$ and $L_1=40$.}
    \label{fig.transverse_motion_cloud} 
\end{figure}

The setup for the transverse motion is shown in Fig.~\ref{fig.motion_scene}(b). In this experiment, a fan controlled by a stepper motor and an Arduino MCU was positioned so that the sensor imaged the middle and top blades. With asynchronous settings of $\alpha=8$ and $L_1=40$, the resulting event point cloud is shown in Fig.~\ref{fig.transverse_motion_cloud}. The rotation of the fan is clearly visible, with the right blade moving from the bottom rows of the sensor to the top rows around columns 50--60 of the imager.

\subsection{Dynamic Depth Data}
Although the proposed asynchronous system provides good tracking of dynamic objects, as shown in the previous section, its output is still dominated by redundant information. For example, the background wall in both experiments was repeatedly reported. Such redundancy could lead to wasted computational power and resources for downstream processing. To address this issue, we present a method to pre-filter peak events into a format more closely resembling that of dynamic vision sensors (DVS). In DVS cameras, only two types of events are generated to indicate positive or negative changes in intensity. Inspired by this concept, we propose changing the output format of the asynchronous LiDAR to \textit{Dynamic Depth} (DD), where events represent only changes in pixel depth. If the depth change exceeds a predefined threshold, a positive or negative DD event is generated depending on the motion polarity.

\begin{figure}[htbp]
    \centering
    \includegraphics[width=\linewidth]{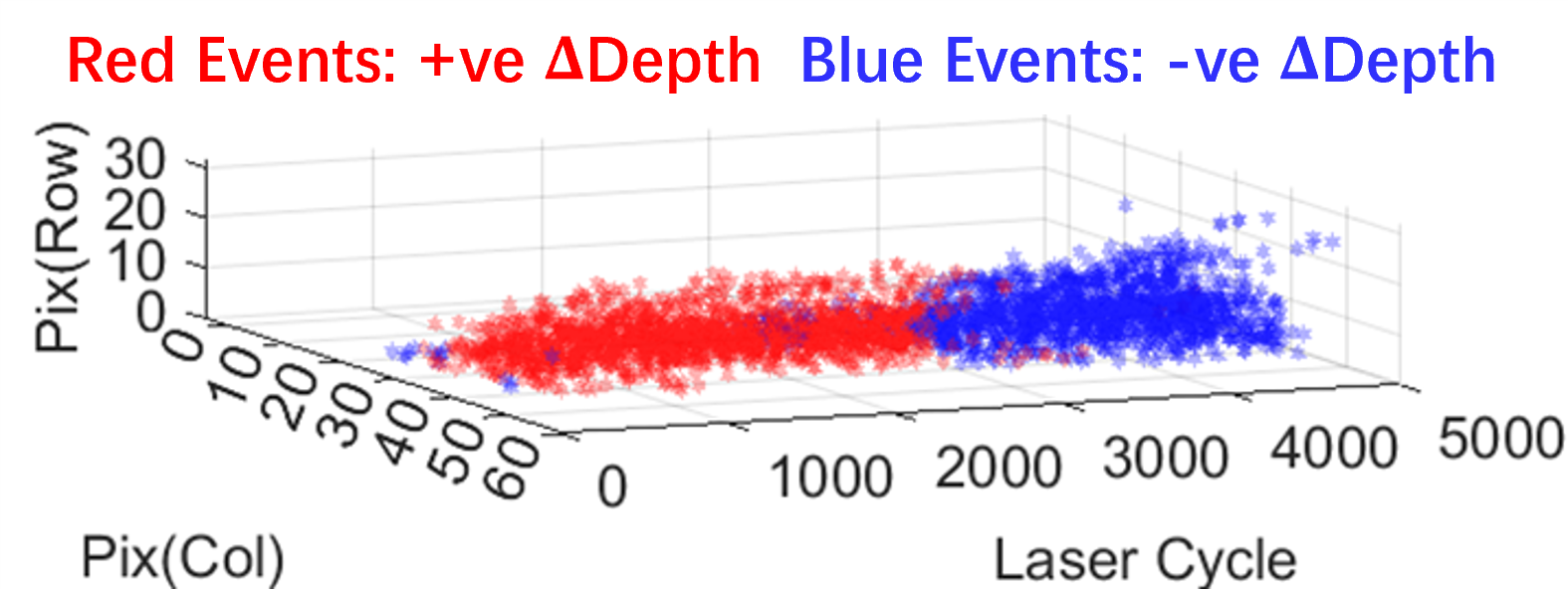}
    \caption{Dynamic Depth data of the radial motion.}
    \label{fig.radial_DD} 
\end{figure}

Based on the peak event data from the radial motion in Fig.~\ref{fig.radial_motion_cloud}, the DD output was obtained by comparing the three-event moving-averaged depth of two adjacent peak events with a threshold of $\pm0.1$~m. The resulting DD data are shown in Fig.~\ref{fig.radial_DD}. The DD events closely track the programmed motion of the car described in Section~\ref{sec.motion}. Since the background wall remained static, no DD events were generated for those pixels.

\begin{figure}[!h]
    \centering
    \includegraphics[width=\linewidth]{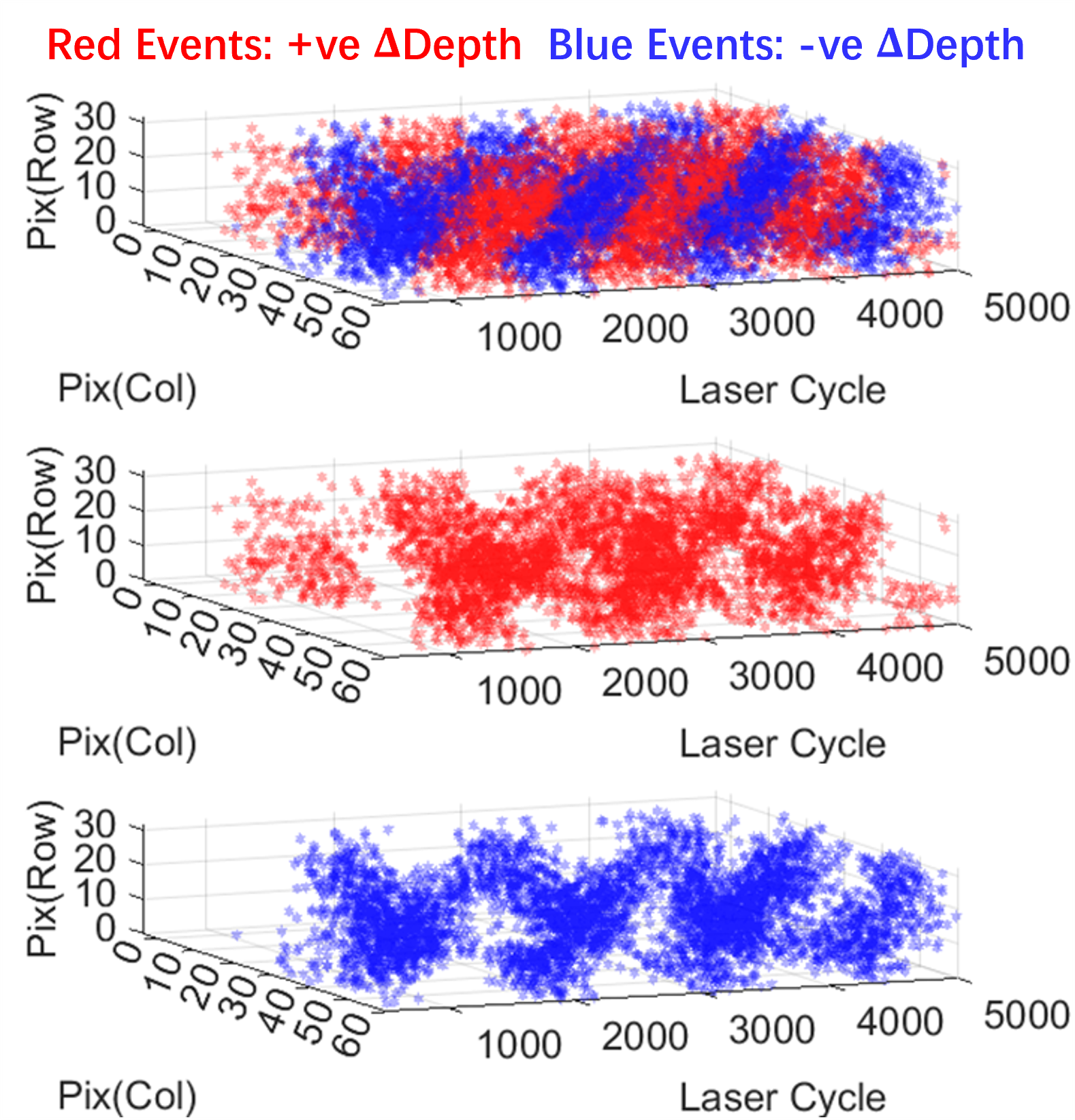}
    \caption{DD data of the transverse motion.}
    \label{fig.transverse_DD} 
\end{figure}

The same procedure was applied to the transverse motion, and the results are shown in Fig.~\ref{fig.transverse_DD}. Similar to the radial case, background pixels (e.g., columns 0--10) did not produce DD events. The rotation of the fan is clearly visible: DD events capture depth transitions from the fan to the background wall and back to the fan. Overall, an event rate compression of over 70\% could be achieved by comparing the number of events in DD output plots and the original peak event plots. This reduction significantly lowers data bandwidth and computational load, improving system efficiency in real-time applications.

\section{Proposed Hardware Implementation\label{Sec.4}}
\subsection{Pixel Architecture}
In this section, we present a potential hardware implementation of the asynchronous operation approach. In the proposed design, SPADs in the top tier are combined into macropixels through pixel binning, with configurations of either $3\times3$ or $4\times4$, depending on the available circuit area in the bottom tier. In the bottom tier, apart from the essential SPAD front-end, TDC, and histogram modules, the remaining area is used to implement a processing element for four macropixels. The histograms of these pixels are multiplexed and alternately transmitted to the processing block, equivalent to $X=4$ for each pixel. The block detects the peak concurrently during the exposure, ensuring that no additional latency is introduced. When a peak event is detected, a REQ handshake signal is sent to read out the AER blocks.

\begin{figure}[htbp]
    \centering
    \includegraphics[width=\linewidth]{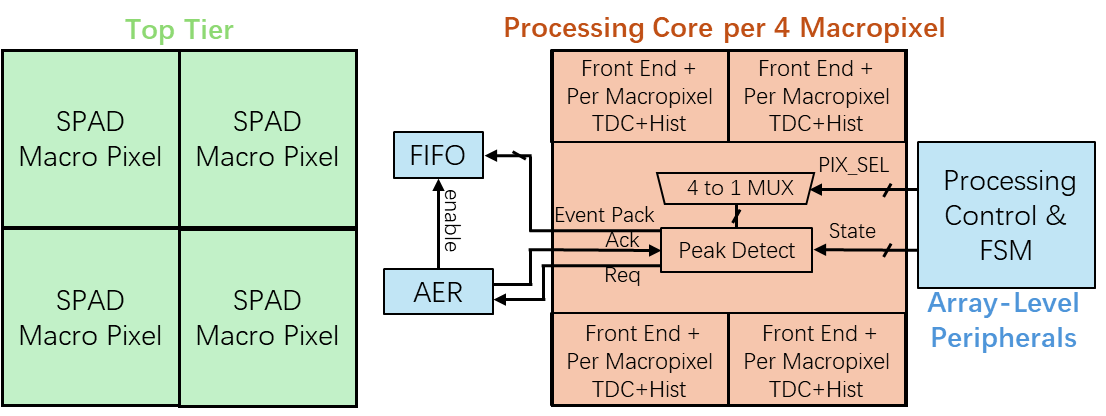}
    \caption{Proposed architecture for each group of four macropixels.}
    \label{fig.architecture} 
\end{figure}

\subsection{FPGA Proof-of-Concept}
\begin{figure}[htbp]
    \centering
    \includegraphics[width=\linewidth]{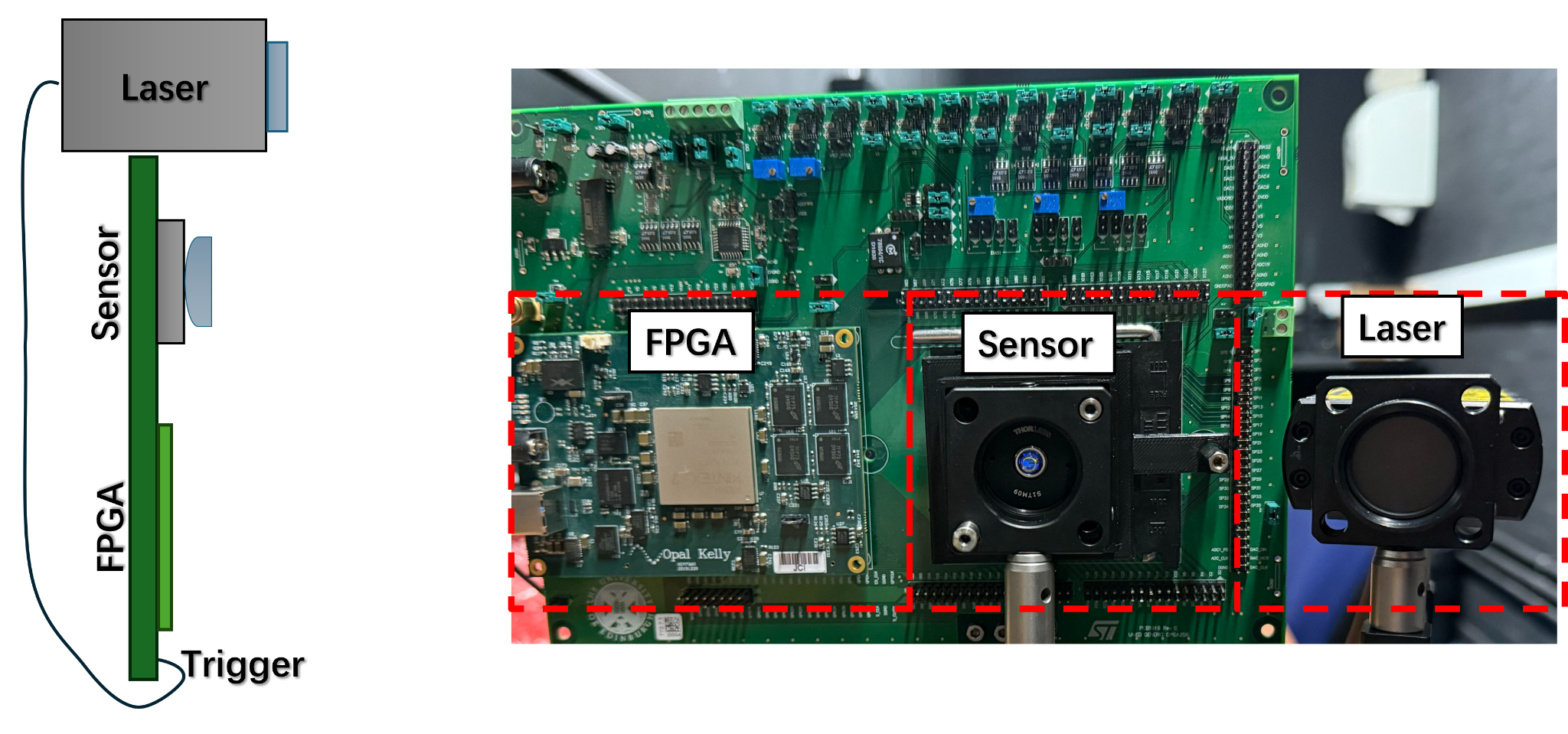}
    \caption{System setup for hardware implementation validation.}
    \label{fig.FPGA system} 
\end{figure}

To validate the feasibility of this architecture, we assembled an electronically scanned LiDAR system based on a $128\times64$ SPAD imager. In this system, SPADs were grouped into 64 subgroups that were multiplexed and directly output to the pads after the front-end circuit, without any on-chip processing \cite{FPGA_sensor}. The remaining modules—including per-macropixel TDCs, histograms, the processing core, and peripheral controls—were implemented on an Opal Kelly XEM7360 FPGA. A photograph of the system is shown in Fig.~\ref{fig.FPGA system}. The laser source used was a Hamamatsu M1030615 773~nm laser head with a pulse width of 120~ps and a repetition rate of 10~MHz.

\begin{figure}[htbp]
    \centering
    \includegraphics[width=\linewidth]{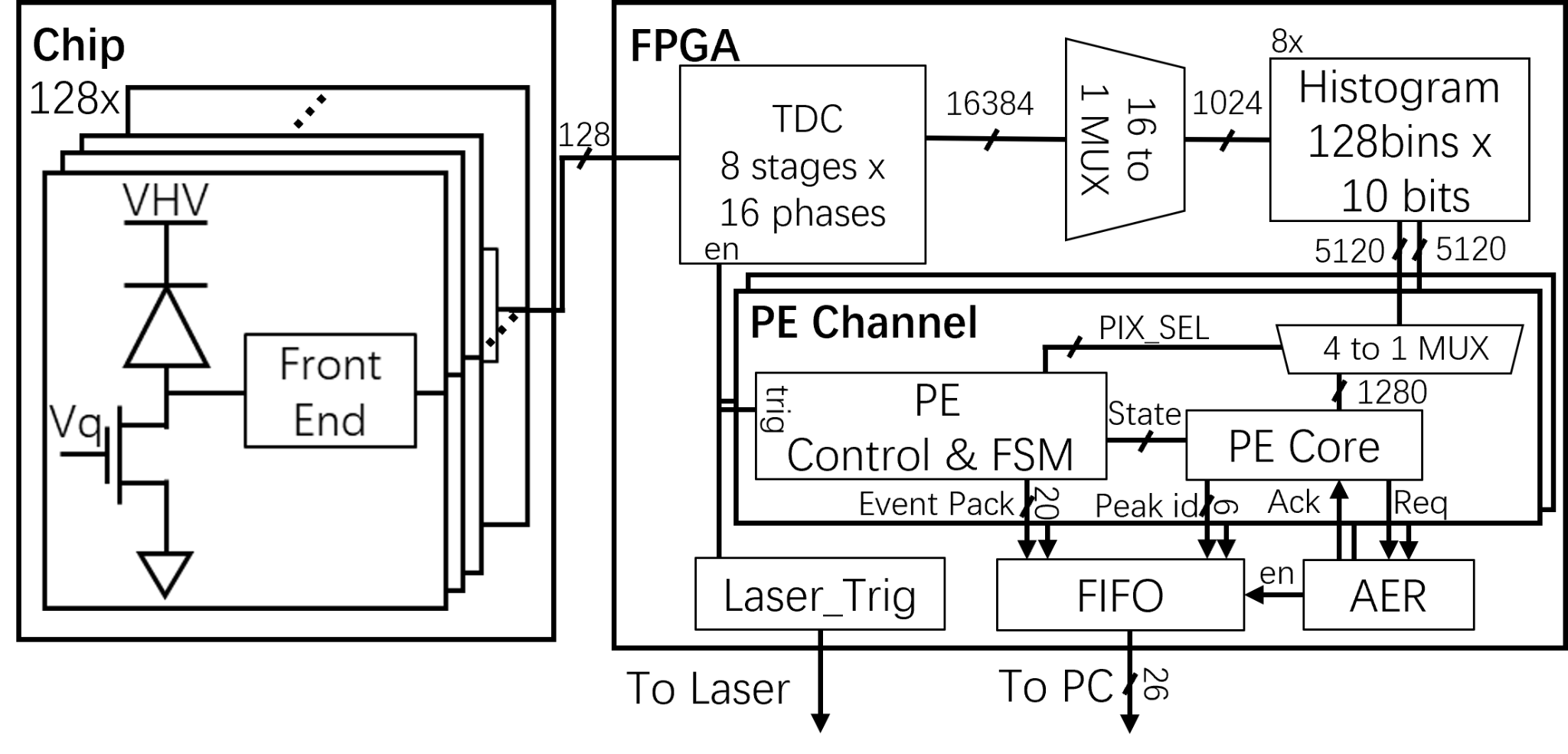}
    \caption{Block diagram of the system modules.}
    \label{fig.FPGA blockdia} 
\end{figure}

\begin{figure}[htbp]
    \centering
    \includegraphics[width=\linewidth]{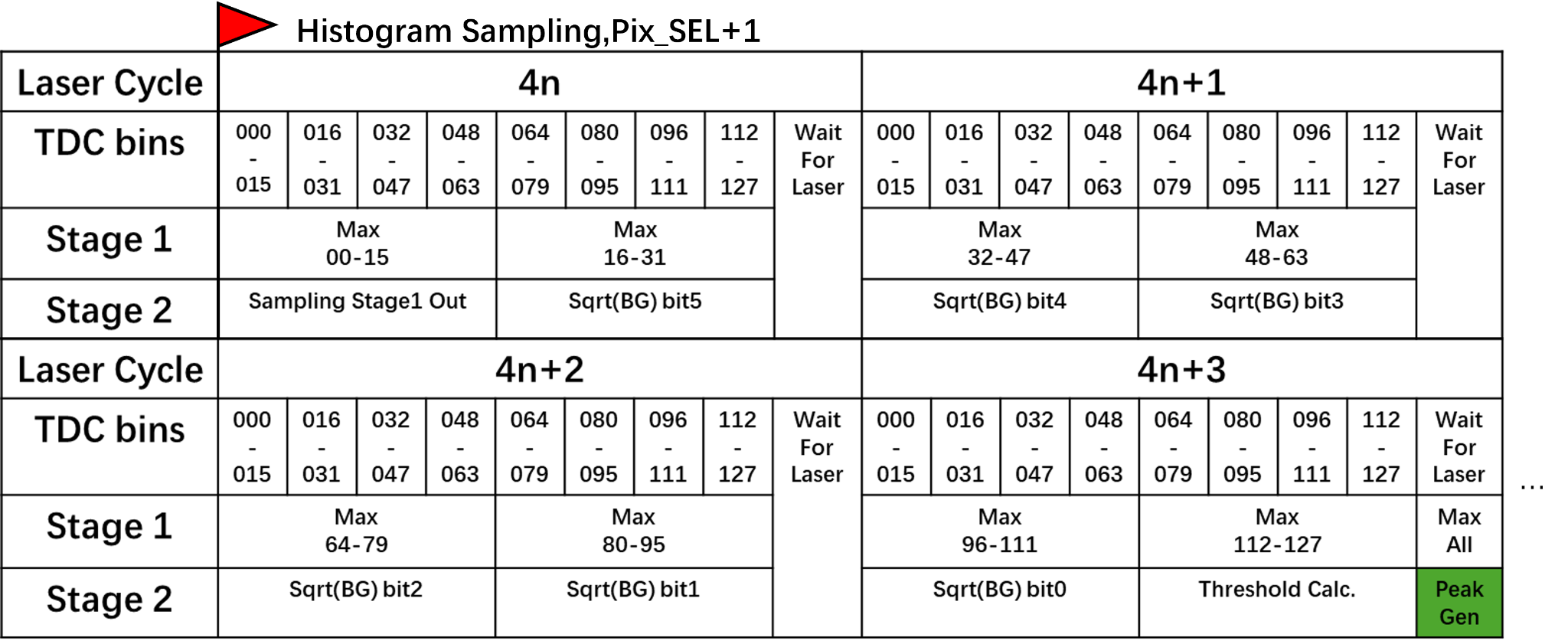}
    \caption{Details of the processing element with two parallel stages.}
    \label{fig.parallel} 
\end{figure}

The block diagram of the circuit modules is shown in Fig.~\ref{fig.FPGA blockdia}. Each SPAD subgroup output was fed into a TDC with 250~ps bin resolution, producing 128-bin, 10-bit histograms. The histograms were then 4-to-1 multiplexed into a shared processing channel, which performed peak finding and thresholding. Each histogram takes four laser cycles to be processed by the processing channel. To improve temporal efficiency, processing was carried out in parallel with ToF histogram accumulation, as illustrated in Fig.~\ref{fig.parallel}.

The algorithm implementation consisted of two stages. In the first stage, the histogram bins were divided into eight groups and compared using an 8-to-1 comparator tree. After comparison of all 128 bins, the maximum bin from each group was compared again to identify the global peak. The second stage calculated the background value, determined the threshold, and compared the peak against this threshold to decide whether to generate a peak event. Computing the background as the average of all 128 bins was found to be time- and resource-intensive. Instead, the background was estimated using the maximum bin of the smaller quadrant of the non-peak half of the histogram. For example, if the peak was located in bins 64--127, the non-peak half was defined as bins 0-–63, and the smaller quadrant was determined by comparing the maxima of bins 0-–31 and bins 32-–63. This approach produced the background value as a byproduct of peak finding, and avoided issues when the signal peak was located near the histogram center.

\begin{figure}[htbp]
    \centering
    \includegraphics[width=\linewidth]{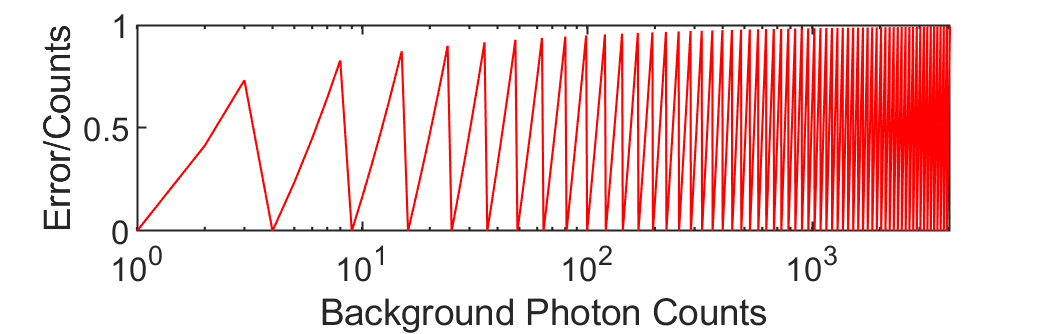}
    \caption{Error of the square-root computation algorithm.}
    \label{fig.sqrt_error} 
\end{figure}

In the second stage, an iterative algorithm was used to compute the square root of the background. The pseudocode is provided in Alg.~\ref{alg.1}. This algorithm maintained an error below one for all inputs and required five clock cycles to compute the square root, as shown in Fig.~\ref{fig.sqrt_error}. Compared with other square-root algorithms used in LiDAR thresholding \cite{HSLiDAR, Park_ISSCC2025}, the proposed method achieved the highest precision and accuracy at the expense of slightly longer computation, which was acceptable as the processing was pipelined, therefore no extra latency was introduced by using this algorithm.

\begin{algorithm}
\caption{Computing the Square Root of the Background Counts.}
\begin{algorithmic}[1]
    \State $temp \gets 0$
    \State $v\_bit \gets 5$
    \State $n \gets 0$
    \State $b \gets 32$
    \State $BG\_temp \gets BG$
    
    \While{$b > 0$}
        \State $temp \gets (n << 1 + b) << v\_bit$
        \State $v\_bit \gets v\_bit - 1$
        
        \If{$BG\_temp \geq temp$}
            \State $n \gets n + b$
            \State $BG\_temp \gets BG\_temp - temp$
        \EndIf
        
        \State $b \gets b >> 1$
    \EndWhile
    
    \State Store $n$ as $sqrt\_BG$
\end{algorithmic}
\label{alg.1}
\end{algorithm}

\subsection{Experiment Result}
\begin{figure}[htbp]
    \centering
    \begin{subfigure}[b]{0.39 \linewidth}
        \centering
        \includegraphics[width=\textwidth]{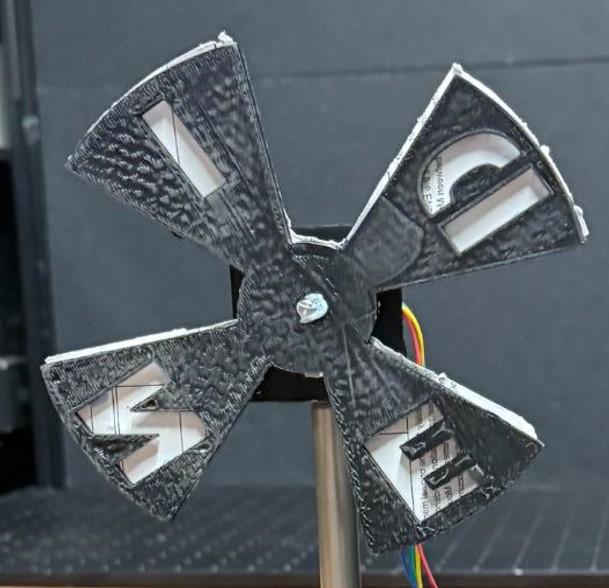}
    \end{subfigure}
    \begin{subfigure}[b]{0.51 \linewidth}
        \centering
        \includegraphics[width=\textwidth]{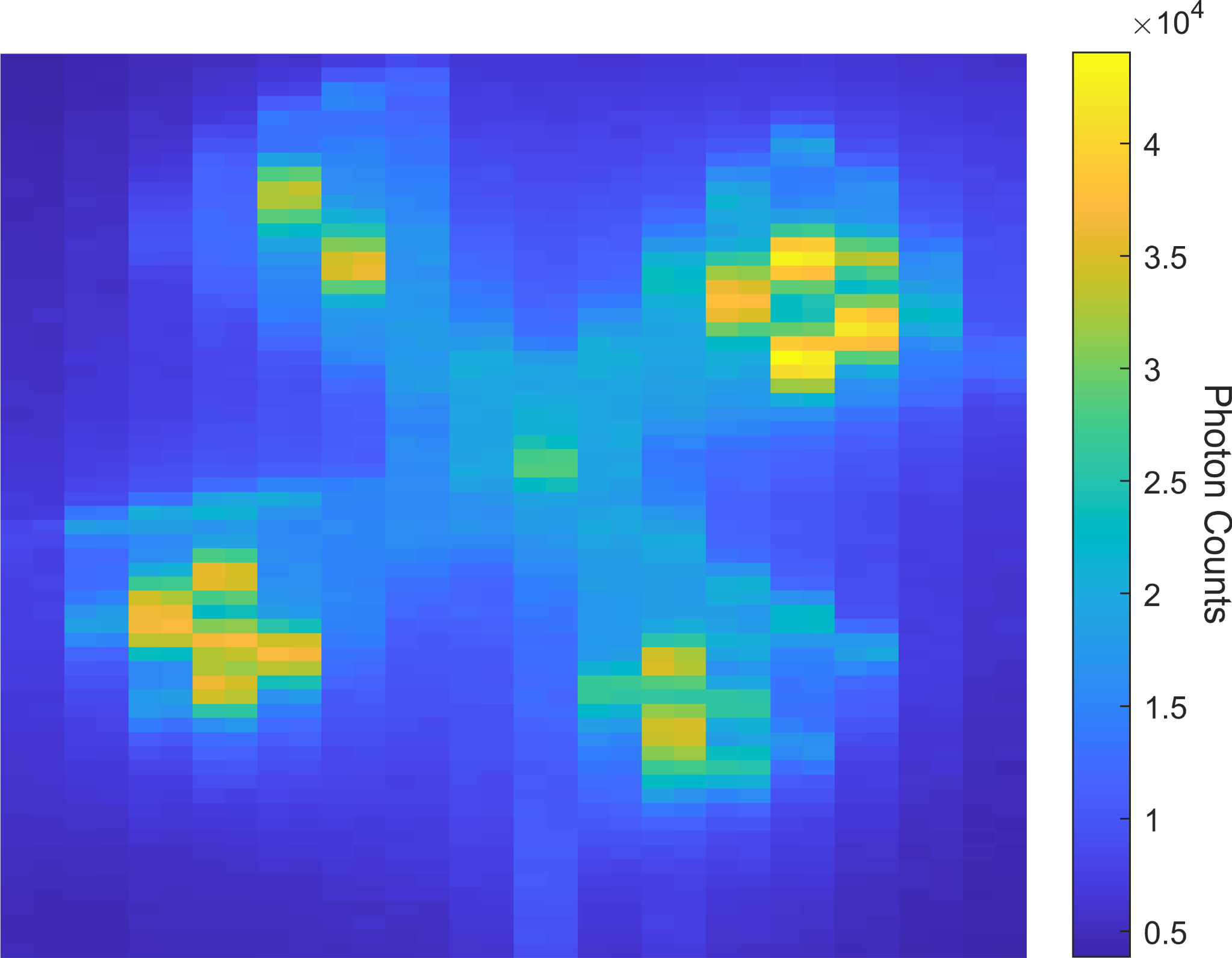}
    \end{subfigure}
    \caption{Imaging scene captured by the RGB camera and the SPAD imager in intensity mode.}
    \label{fig.intensity_FPGA}
\end{figure}

Similar to the PC-based system, the FPGA-based system was tested in both static and dynamic experiments. The same fan setup shown in Fig.~\ref{fig.motion_scene}(b) was used. The intensity images acquired by the RGB camera and the SPAD imager are shown in Fig.~\ref{fig.intensity_FPGA}.

\begin{figure}[htbp]
    \centering
    \includegraphics[width=\linewidth]{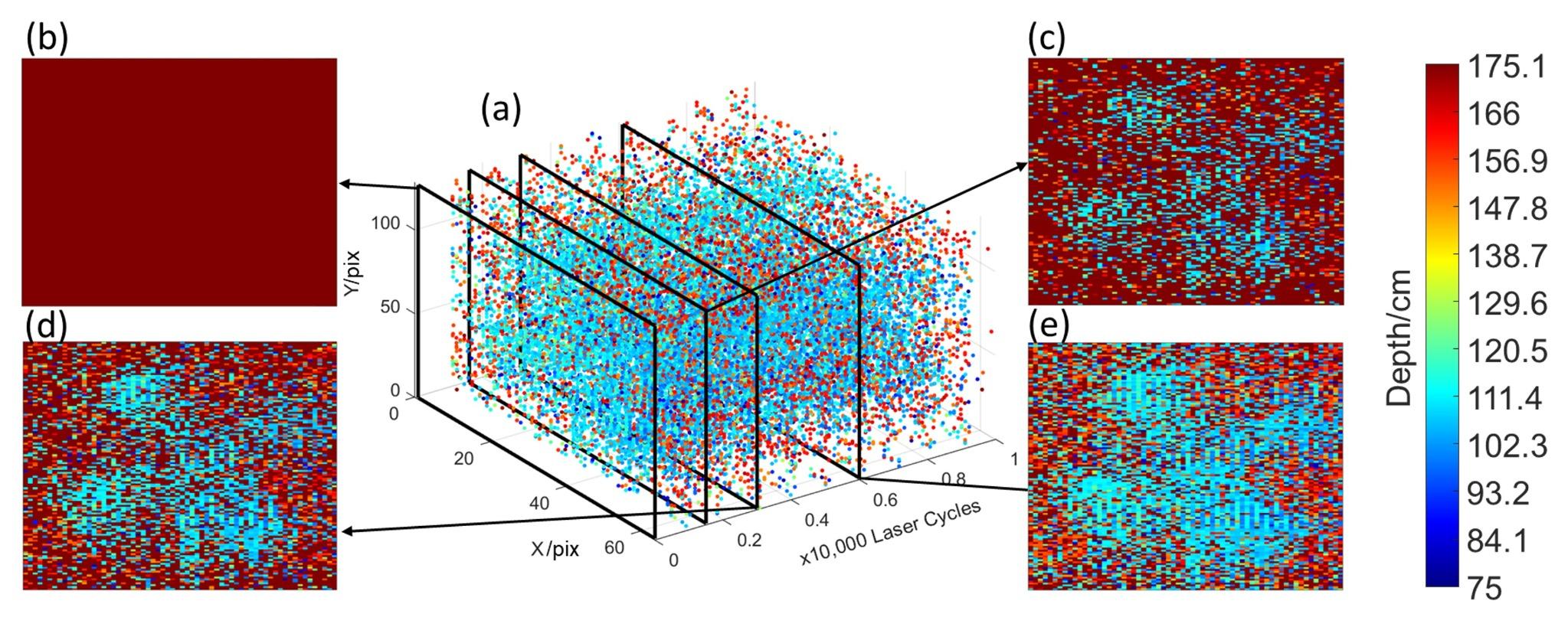}
    \caption{Event point cloud of the static fan.}
    \label{fig.static_FPGA} 
\end{figure}

\begin{figure}[htbp]
    \centering
    \includegraphics[width=\linewidth]{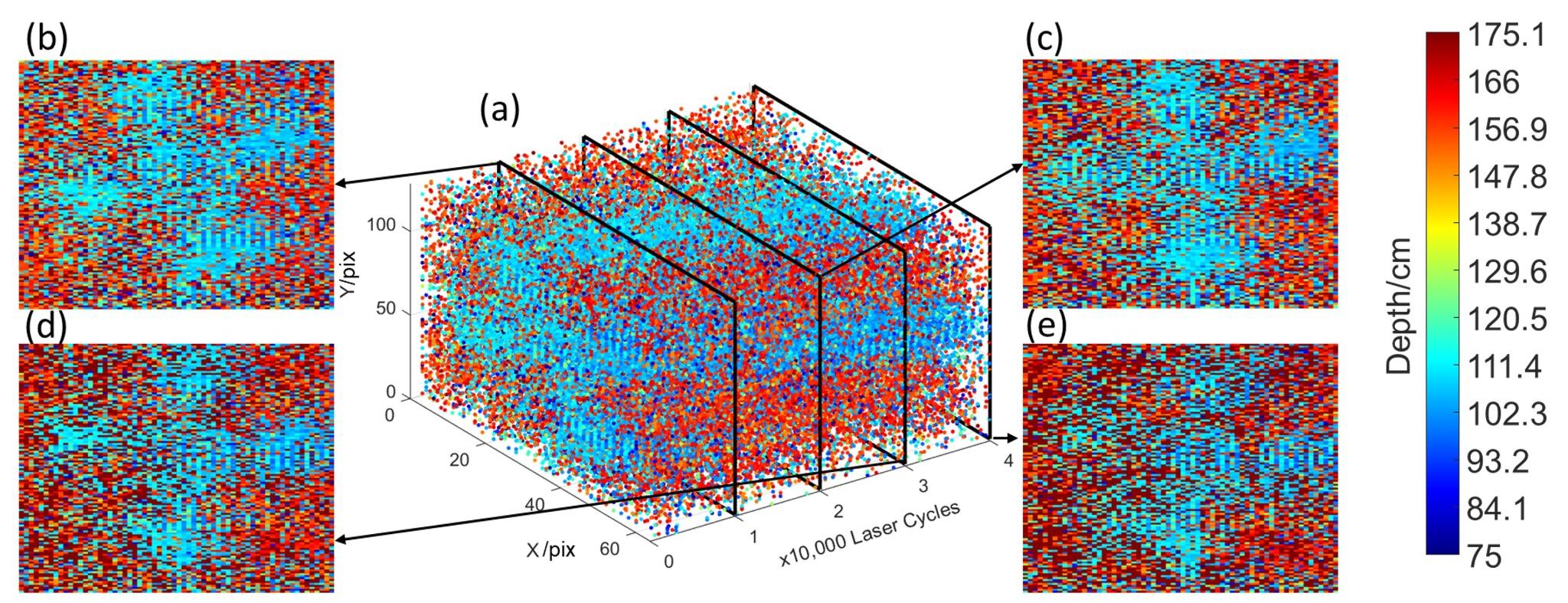}
    \caption{Event point cloud of the fan rotating clockwise at 120 rpm.}
    \label{fig.dynamic_FPGA} 
\end{figure}

The event point clouds generated by the FPGA-based system for static and dynamic scenes are shown in Fig.~\ref{fig.static_FPGA} and Fig.~\ref{fig.dynamic_FPGA}, respectively. For the static scene, similar to the offline-processed system, the peak events delineated the fan from the background. From the four depth-map slices, it was observed that pixels began reporting peak events at different times depending on their SNR. For the dynamic experiment, the fan was rotated clockwise at 120~rpm. The resulting event point cloud shows that the system successfully tracked the transverse motion without motion blur.

\section{Discussion and Conclusion\label{Sec.5}}

\begin{table*}[htbp]
\centering
\caption{Comparison with representative LiDAR systems with dynamic or asynchronous outputs}
\begin{tabular}{c|c|c|c|c|c|c}
\hline
 & \textbf{This work\textsuperscript{(1)}} & \textbf{\cite{Park_ISSCC2025}} & \textbf{\cite{Gomez_SPADTTFS}} & \textbf{\cite{HSLiDAR}} & \textbf{\cite{Yao:24}} & \textbf{\cite{Kitichotkul:25}} \\
\hline
\textbf{Data Type} & Peak Events/DD & Peak Events & TTFS/ Dynamic Vision & DD & DD & Speed \\
\textbf{Data Format} & Async & Async Frames & Async & Frame Based & Frame Based & Frame Based \\
\textbf{Processing} & In Pixel & In Pixel & In Pixel/FPGA & Col-Parallel & Post-Processing & Post-Processing \\
\textbf{Frame rate/fps} & -- & 5--250\textsuperscript{(2)} & -- & 50 & N/A & 50 \\
\hline
\multicolumn{7}{c}{\textbf{LiDAR Hardware Comparison}} \\
\hline
\textbf{LiDAR Circuitry} & On-Chip & On-Chip & Off-Chip & On-Chip &  &  \\
\textbf{Technode} & 40nm & 90nm & 110nm & 40nm &  &  \\
\textbf{Array Size} & 64$\times$32 & 40$\times$90 & 64$\times$64 & 64$\times$32 & (3) & (3) \\
\textbf{Power Consumption/ mW} & 70+55\textsuperscript{(4)} & 84 & -- & 70 &  &  \\
\textbf{Latency/ $\upmu $s} & \textbf{2.4\textsuperscript{(5)}} & 4000 & 128 & 20000 &  &  \\
\hline
\end{tabular}
\label{tab.fom}
\begin{tablenotes}
\scriptsize
\item[1] (1) Estimated based on \cite{HSLiDAR}, assuming the same Frontend, TDC, histogram circuitry.  \\
\item[2] (2) Frame rate depends on pixel SNR (dynamic).  \\
\item[3] (3) No on-chip implementation possible \\
\item[4] (4) 70 mW for sensor operation based on \cite{HSLiDAR}, and 55 mW for processing estimated from FPGA report, assuming 10\% power from FPGA. \\
\item[5] (5) When $L_1=40$ and 10 MHz Laser rate are used.
\end{tablenotes}
\end{table*}
\vspace{-0.5em}

In this work, we introduced a fully asynchronous peak detection approach for SPAD-based dToF flash LiDAR systems, where each pixel independently determines when a sufficient signal-to-noise ratio is reached and reports events without global synchronization. Through theoretical analysis, simulations, and experimental validation, we showed that this method reduces latency, mitigates motion blur, and improves effective frame rate while maintaining accuracy under both static and dynamic conditions. The framework also supports reflectivity reconstruction and a DVS-inspired event representation, broadening its applicability to real-time 3D sensing. We validated the framework using two hardware setups: an offline system that processes the LiDAR data with an emulated asynchronous processor on a PC, and a real-time FPGA-based prototype. Both demonstrated robust depth estimation, reflectivity recovery, and dynamic event-based representation, with tunable performance through simple hyperparameters such as $L_1$ and $\alpha$. 

The comparison between the proposed work to some selected state-of-the-art related work is stated in Table.~\ref{tab.fom}. Compared to other computational methods for velocity or DD extraction \cite{Kitichotkul:25, Yao:24}, our approach provides a simpler, hardware-friendly solution suitable for SPAD-SoC integration, reducing cost and latency while enhancing scalability. Compared to the other hardware based architectures, our proposed methods could provide the best latency with acceptable power increment, which is the key metric for motion tracking.

Overall, looking ahead, the proposed architecture opens opportunities for event-driven LiDAR systems that align with neuromorphic vision and spiking neural networks. Future directions include on-chip integration of processing elements, refined event encoding strategies, and adaptation for outdoor and high-dynamic-range environments. By advancing asynchronous depth sensing, this work establishes a foundation for next-generation LiDAR systems that combine high temporal resolution, low latency, and energy efficiency.

\section*{Acknowledgments}
The authors thank Germ{\'a}n Mora-Mart{\'\i}n of Edinburgh for helping to provide data used in the preliminary modeling stage of this work. They also thank Ahmet T. Erdogan of the University of Edinburgh for designing the sensor used in the testing system, and ST Microelectronics for fabricating the chip.

\bibliographystyle{IEEEtran}
\bibliography{CITES.bib}
\vfill

\end{document}